\documentclass[11pt]{iopart}
\usepackage{amsfonts}
\usepackage{graphicx}
\usepackage{iopams}
\eqnobysec
\usepackage{tikz}
\newcommand{\ot}[0]{\otimes}
\newcommand{\bei}{\begin{itemize}}
\newcommand{\eei}{\end{itemize}}

\newcommand{\proj}[1]{\ket{#1}\!\bra{#1}}
\newcommand{\braket}[2]{\langle{#1}|{#2}\rangle}

\newcommand{\HA}{\mathcal{H}_A}
\newcommand{\HB}{\mathcal{H}_B}
\newcommand{\HAB}{\mathcal{H}_{AB}}
\newcommand{\HT}{\mathcal{H}_{\rm total}}
\newcommand{\ra}{{\, \rightarrow\, }}

\newcommand{{\Cd}}{{\mathbb{C}^3}}

\def\<{\langle}
\def\>{\rangle}
\newcommand{\ket}[1]{|#1\rangle}
\newcommand{\bra}[1]{\langle#1|}

\newtheorem{Theorem}{Theorem}[section]
\newtheorem{Proposition}{Proposition}[section]
\newtheorem{Definition}{Definition}[section]
\newtheorem{Example}{Example}[section]
\newtheorem{Remark}{Remark}[section]
\newtheorem{Cor}{Corollary}[section]
\newtheorem{Lemma}{Lemma}[section]
\begin{document}

%\topical[Entanglement witnesses:  theory and applications]{Entanglement witnesses:  theory and applications}

%\topical
%\title[

\title[Entanglement witnesses: construction, analysis  and classification]{Entanglement witnesses: construction, analysis  and classification}

\author{Dariusz  Chru\'sci\'nski and
Gniewomir Sarbicki}

\address{Institute of Physics, Nicolaus Copernicus University,\\
Grudzi\c{a}dzka 5/7, 87--100 Toru\'n, Poland}
\ead{darch@fizyka.umk.pl; gniewko@fizyka.umk.pl}

%\maketitle

\begin{abstract}

From the physical point of view entanglement witnesses define a universal tool for analysis and classification of quantum entangled states. From the mathematical point of view they provide highly nontrivial generalization of positive operators and they find elegant correspondence with the theory of positive maps in matrix algebras. We concentrate on theoretical analysis of various important notions like (in)decomposability, atomicity, optimality, extremality and exposedness. Several methods of construction are provided as well. Our discussion is illustrated by many examples enabling the reader to see the intricate structure of these objects. It is shown that the theory of entanglement witnesses finds elegant geometric formulation in terms of convex cones and related geometric structures.

\end{abstract}

%\tableofcontents

%\maketitle
\section{Introduction}

Quantum entanglement is one of the key features of quantum physics.
The interest on quantum entanglement has dramatically
increased during the last two decades due to the emerging field
of quantum information theory. It turns out that quantum
entangled states may be used as basic resources in quantum
information processing and communication. The prominent
examples are quantum cryptography, quantum teleportation,
quantum error correction codes, and quantum computation.
It is therefore clear that there is a considerable interest in efficient theoretical
and experimental methods of entanglement analysis, classification and detection.  There are several excellent review articles
(see for example \cite{HHHH} by Horodecki family and \cite{Guhne} by G\"uhne and Toth)  and books \cite{QIT,KAROL,VEDRAL} dealing with quantum entanglement and quantum information.

One of the essential problems in quantum entanglement theory is the classification of states of composite quantum systems. In particular it is of primary importance to test whether a given quantum state is separable or entangled. For low dimensional systems (qubit-qubit or qubit-qitrit) there exists a necessary and sufficient condition for separability for bipartite systems -- the celebrated Peres-Horodecki criterion based on positivity of partial transposition.  However, for higher dimensional systems, or more number of parties, there is no single universal separability condition. There are several theoretical and experimental tools enabling one to analyze and detect quantum entanglement. Many authors contributed to a long list of various separability criteria and detection methods \cite{HHHH,Guhne}.

The most general approach to characterize quantum entanglement uses a notion of an entanglement witness. The term {\em entanglement witness} for operators detecting quantum entangled states was introduced by Terhal \cite{EW2}. One of the big advantages of entanglement witnesses is that they provide an economic method of detection which does need the full information about the quantum state. Such information is usually obtained by the full state tomography. Here one uses only the information about the mean value of some observable in a given quantum state. Remarkably, it turns out that any entangled state can be detected by some entanglement witness and hence the knowledge of witnesses enables us to perform full classification of states of composite quantum systems. Interestingly, entanglement witnesses are deeply connected to a theory of positive maps in operator algebras.  Positive maps play an important role both in physics and mathematics providing generalization of $*$-homomorphisms, Jordan homomorphisms and conditional expectations. In the algebraic approach to quantum physics \cite{Haag} normalized
positive maps define  affine mappings between sets of states of $\mathbb{C}^*$-algebras.

The review is organized as follows:  section \ref{S-STATES} discusses the structure of states for bipartite quantum systems. We introduce the notion of Schmidt number for an arbitrary bipartite positive operator and perform classification of quantum states. Special role of separable states, states positive under partial transposition (PPT), and so called edge states is emphasized. Section \ref{S-EW} provides basic definitions and properties of entanglement witnesses. We introduce the notion of a $k$-Schmidt witnesses, (in)decomposable and atomic witnesses and finally provide a general representation of an indecomposable entanglement witness based on the properties of edge states. Section \ref{MAPS} analyzes positive maps in operator algebras and discusses well known correspondence between linear maps and bipartite operators. Hence, all properties of witnesses may be analyzed in the language of maps and vice versa. Section \ref{SEP-EW} shows how well known separability criteria may be used for construction natural entanglement witnesses. In particular it is shows how Bell inequalities may be connected to appropriate entanglement witnesses.
Section \ref{S-OPTIMAL} introduces important notions of optimal, extremal and exposed witnesses.

Sections \ref{S-DIAG} and \ref{S-OPT} illustrate introduced theoretical concepts with several well known examples:  section \ref{S-DIAG} analyzes the structure of so called diagonal-type entanglement witnesses for qudit-qudit systems. In particular it discusses witnesses constructed via generalization of the celebrated Choi positive indecomposable map. Section \ref{S-OPT} discusses generalization of another `classical' map constructed by Robertson. Remarkably, entanglement witnesses constructed this way are optimal and hence provide the strongest theoretical tool for detecting quantum entangled states.
Section \ref{CIRCULANT} provides important construction of witnesses based on so called circulant decomposition of the Hilbert space corresponding to composite system of two qudits. This class contains for example well known family of Bell diagonal witnesses. Section \ref{K} shows how to construct $k$-Schmidt witness and it is based on the spectral decomposition of the corresponding witness. This construction is up to now the most general one known in the literature. In Section \ref{S-MULTI} we show how to generalize the analysis of quantum states and entanglement witnesses to multipartite scenario.
Section \ref{S-CONES} may be treated as a completion of the paper. It shows how the geometry of convex cones enables one to present the structure and properties of entanglement witnesses in a unified elegant  geometric way.

\section*{Basic notation}

Let us introduce a basic notation used throughout the paper: we denote by  $\mathfrak{L}(\mathcal{H})$ a vector space of linear operators acting on a finite dimensional Hilbert space $\mathcal{H}$. Fixing a basis in $\mathcal{H}$ one may identify $\mathfrak{L}(\mathcal{H})$ with the space $M_n(\mathbb{C})$ of $n \times n$ complex matrices with $n = {\rm dim}\, \mathcal{H}$. One endows $\mathfrak{L}(\mathcal{H})$ with the Hilbert-Schmidt inner product $\<A|B\>_{\rm HS} = \tr (A^\dagger B)$. If $\{e_1,\ldots,e_n\}$ denotes an orthonormal basis in $\mathcal{H}$, then
\begin{equation*}\label{}
  \<A|B\>_{\rm HS} = \sum_{i=1}^n \< Ae_i|B e_i\> \ , \nonumber
\end{equation*}
where $\<\psi|\phi\>$ denotes an inner product in $\mathcal{H}$. This product gives rise to the Hilbert-Schmidt norm
\begin{equation*}\label{}
  || A ||_{\rm HS} = \sqrt{ \<A|A\>_{\rm HS} } = \sqrt{ \tr A^\dagger A } \ . \nonumber
\end{equation*}
Apart from  $|| A ||_{\rm HS}$ the space $\mathfrak{L}(\mathcal{H})$ is endowed with  the standard operator norm
\begin{equation}\label{}
  || A || = \sup_\psi \frac{ ||A\psi||}{||\psi||} \ , \nonumber
\end{equation}
and the trace norm
\begin{equation}\label{}
  || A ||_1  =  {\rm tr} |A| = {\rm tr} \sqrt{A^\dagger A}  \ . \nonumber
\end{equation}
If $\lambda_1^2 \geq \lambda_2^2 \geq \ldots \geq \lambda_n^2$ are eigenvalues of $A^\dagger A$, then
\begin{equation*}\label{}
  || A ||_{\rm HS} = \sqrt{ \lambda_1^2 + \ldots + \lambda_n^2}\ , \ \  ||A|| = |\lambda_1|\ , \ \  ||A||_1 = |\lambda_1| + \ldots + |\lambda_n|\ . \nonumber
\end{equation*}
Denote by $\mathfrak{L}_+(\mathcal{H})$ a subspace of positive operators in $\mathfrak{L}(\mathcal{H})$. Note that $\mathfrak{L}_+(\mathcal{H})$ is no longer a vector space but it defines a  convex cone in $\mathfrak{L}(\mathcal{H})$ (see Section \ref{S-CONES} for details about convex cones).  Finally, mixed states represented by density operators give rise to compact convex set
\begin{equation*}\label{}
  \mathfrak{S}(\mathcal{H}) = \{ \rho \in \mathfrak{L}_+(\mathcal{H})\, |\, \tr \rho =1 \}  \ , \nonumber
\end{equation*}
i.e. a set of normalized positive operators. It is clear that $A \in \mathfrak{L}_+(\mathcal{H})$ if and only if $||A||_1 = \tr A$.

\section{States of bipartite quantum systems}  \label{S-STATES}

Consider a Hilbert space $\mathcal{H}_{AB} = \HA \otimes \HB$ such that dimensions ${\rm dim}\HA=d_A$ and ${\rm dim}\HB = d_B$  are finite. Let us denote
$D:=d_Ad_B$ and $d:=\min\{d_A,d_B\}$. Now, for any  vector $\psi \in \mathcal{H}_{AB}$ one has the corresponding Schmidt decomposition
\begin{equation}\label{}
    \psi = \sum_{k=1}^r \mu_k\, e_k \otimes f_k\ ,
\end{equation}
where $\mu_k > 0$ and  $\{e_i\}$, $\{f_j\}$ are two families of orthogonal normalized vectors in $\HA$ and $\HB$, respectively. If $\psi$ is normalized, i.e. it corresponds to a pure state of a composite system living in $\HAB$,  then $s_k(\psi) := \mu_k$ are called  Schmidt coefficients of $\psi$ and they satisfy $\sum_{k=1}^r [s_k(\psi)]^2 = 1$. One calls the number `$r$' the Schmidt rank ${\rm SR}(\psi)$ of $\psi$.  It is clear that $1 \leq {\rm SR}(\psi) \leq d$.

\begin{Definition}
A vector $\psi \in \HAB$ is separable iff $\psi = \psi_A \otimes \phi_B$, that is, ${\rm SR}(\psi)=1$,   and entangled otherwise. Vector $\psi$ is called maximally entangled iff $r=d$ and $s_k(\psi) = 1/\sqrt{d}$ for $k=1,\ldots,d$.
\end{Definition}
For example if $\HA=\HB=\mathcal{H}$ and  $\{e_1,\ldots,e_d\}$ is an arbitrary orthonormal basis in $\mathcal{H}$, then
\begin{equation}\label{MAX}
  \psi^+_d = \frac{1}{\sqrt{d}} \sum_{k=1}^d e_k \otimes e_k\ ,
\end{equation}
defines a  "canonical" maximally entangled state in $\mathcal{H}\otimes \mathcal{H}$.
Note that $\HAB$ and the vector space $\mathcal{L}(\HA,\HB)$ of linear operators $F : \HA \rightarrow \HB$ have the same dimension and hence they are isomorphic. Fixing an arbitrary orthonormal basis $\{e_1,\ldots,e_{d_A}\}$ in $\mathcal{H}_A$ let us introduce an isomorphism
${j} : \mathcal{L}(\HA,\HB) \rightarrow \HAB$ by
\begin{equation}\label{X}
  j(F) = \sqrt{d_A} (\mathbb{I}_A \otimes F) \psi^+_{d_A} = \sum_{k=1}^{d_A} e_k \otimes F e_k\ .
\end{equation}
The inverse map $F = j^{-1}(\psi)$ reads as follows: for any vector $\psi = \sum_{k=1}^{d_A} e_k \otimes y_k$, with $y_k \in \HB$, one defines $F e_k = y_k$.
%It is clear that the map $j$ depends upon the basis $\{e_k\}$.
Recall that $\mathcal{L}(\HA,\HB)$ is equipped with a natural inner product
\begin{equation}\label{}
  (F_1,F_2) = {\rm Tr}(F_1^\dagger F_2) \ .
\end{equation}
\begin{Proposition}  \label{PR-j}
The map ${j} : \mathcal{L}(\HA,\HB) \rightarrow \HAB$ provides an isometric isomorphism of two Hilbert spaces
$\mathcal{L}(\HA,\HB)$ and $\HAB$, that is,
\begin{equation}\label{}
  (F_1,F_2) = \< j(F_1)|j(F_2)\> \  . %= \<\psi_1|\psi_2\>\ .
\end{equation}
In particular $||F|| = ||\psi||$.
\end{Proposition}
It is therefore clear that operators $F \in \mathcal{L}(\HA,\HB)$ satisfying $\tr (F^\dagger F)=1$ correspond to normalized vectors $\psi = j(F)$ in $\HAB$.
Moreover, one has the following

\begin{Proposition} Let $F \in \mathcal{L}(\HA,\HB)$ and  $\psi = j(F)$, then
%\begin{equation}\label{}
 $ {\rm Rank}(F) = {\rm SR}(\psi)$.
%\end{equation}
\end{Proposition}
It is clear that $\psi = j(F)$ does depend upon the basis $\{e_k\}$. However the Schmidt rank of $\psi$  does not.

% Let $\mathfrak{T}(\mathcal{H})$ denotes the trace-class operators in $\mathcal{H}$, i.e. $A : \mathcal{H} \rightarrow \mathcal{H}$ belongs to %$\mathfrak{T}(\mathcal{H})$ if ${\rm Tr}\, |A| < \infty$. Clearly, if $\mathcal{H}$ is finite dimensional, then $\mathfrak{T}(\mathcal{H}) = %\mathcal{L}(\mathcal{H},\mathcal{H})$. A linear space $\mathfrak{T}(\mathcal{H})$ equipped with the trace norm $||A||_1 =  {\rm Tr} |A|$ defines a Banach %space. Let $\mathfrak{T}_+(\mathcal{H})$ denotes a subset of positive operators in $\mathfrak{T}(\mathcal{H})$ (actually, it is convex set cone \ref{CONES}). %The space of mixed states (density operators) in $\mathcal{H}$  is defined as follows

%One has the following
\begin{Definition}[\cite{WERNER}]
A positive operator $ X \in \mathfrak{L}_+(\HAB)$ is separable iff
\begin{equation}\label{}
  X = \sum_k A_k \otimes B_k \ ,
\end{equation}
where $A_k \in \mathfrak{L}_+(\HA)$ and $B_k \in \mathfrak{L}_+(\HB)$.
\end{Definition}
Usually, this definition is formulated in terms of density operators   living in a Hilbert space $\mathcal{H}_{AB}$
\begin{equation}\label{}
   \mathfrak{S}(\mathcal{H}_{AB}) = \{ \rho \in \mathfrak{L}_+(\mathcal{H}_{AB})\ | \  \tr \rho = 1 \} \ .
\end{equation}
A density operator  $\rho \in  \mathfrak{S}(\mathcal{H}_{AB})$ is separable if
\begin{equation}\label{}
  \rho = \sum_k p_k \, \rho^A_k \otimes \rho^B_k \ ,
\end{equation}
where $\rho^A_k \in \mathfrak{S}(\HA)$, $\rho^B_k \in \mathfrak{S}(\HB)$, and $p_k$ is a probability distribution. This definition says that a separable state is defined as a convex combination of product states $\rho^A \otimes \rho^B$.

\begin{Definition}
A Schmidt number ${\rm SN}(X)$ of $X \in \mathfrak{L}_+(\HAB)$ is defined by
\begin{equation}\label{SN-rho}
    {\rm SN}(X) = \min_{\psi_k}\, \left\{ \,
    \max_{k}\, {\rm SR}(\psi_k)\, \right\}  \ ,
\end{equation}
where the minimum is taken over all possible decompositions
\begin{equation}\label{}
    X = \sum_k \, |\psi_k\>\<\psi_k|\ ,
\end{equation}
with  $\psi_k$ being (unnormalized) vectors from $ \HAB$.
\end{Definition}
This definition was originally provided for density operators form $\mathfrak{S}(\HAB)$ by  Horodecki and Terhal \cite{Pawel}. However, the Schmidt number is well defined for any positive operators from $\mathfrak{L}_+(\HAB)$. Note, that if $X = |\psi\>\<\psi|$, then it is clear that
${\rm SN}(\rho) = {\rm SR}(\psi)$ and hence the above definition reproduces the definition of the Schmidt rank of a vector $\psi \in \HAB$.
%\begin{Definition}
%A Schmidt number ${\rm SN}(\rho)$ of $\rho \in \mathfrak{S}(\HAB)$ is defined by
%\begin{equation}\label{SN-rho}
%    {\rm SN}(\rho) = \min_{p_k,\psi_k}\, \left\{ \,
%    \max_{k}\, {\rm SR}(\psi_k)\, \right\}  \ ,
%\end{equation}
%where the minimum is taken over all possible pure states
%decompositions
%\begin{equation}\label{}
%    \rho = \sum_k \, p_k\, |\psi_k\>\<\psi_k|\ ,
%\end{equation}
%with $p_k\geq 0$, $\sum_k\, p_k =1$ and $\psi_k$ are normalizeds
%vectors in $\HAB$.
%\end{Definition}
%
%\begin{Remark} If $\psi \in \HAB$ represents a pure state and $\rho = |\psi\>\<\psi|$, then it is clear that
%$$   {\rm SN}(\rho) = {\rm SR}(\psi)\ , $$
%that is, the above definition reproduces the definition of the Schmidt rank for pure states.
%\end{Remark}
Let us introduce the following family of convex cones (see Section \ref{S-CONES} for appropriate definitions)
\begin{equation}\label{}
    \mathfrak{L}_k = \{\, X \in \mathfrak{L}_+(\HAB)\ |\
    \mathrm{SN}(X) \leq k\, \}  \ .
\end{equation}
One has the following chain of inclusions
\begin{equation}\label{V-k}
\mathfrak{L}_1  \subset \ldots \subset  \mathfrak{L}_d = \mathfrak{L}_+(\HAB)\ .
\end{equation}
%Clearly, $\mathfrak{L}_1$ is a convex cone of separable positive operators
%and $\mathfrak{L}_d \smallsetminus  \mathfrak{L}_1$ stands for a set of entangled positive operators.
Note, that defining
%\begin{equation}\label{}
 $   \mathfrak{S}_k = \mathfrak{L}_k \cap \mathfrak{S}(\HAB)$
%\end{equation}
one finds
\begin{equation}\label{V-k}
\mathfrak{S}_1  \subset \ldots \subset  \mathfrak{S}_d = \mathfrak{S}(\HAB)\ .
\end{equation}
Clearly, $\mathfrak{S}_1$ is a convex subset of separable  states
and $\mathfrak{S}_d \smallsetminus  \mathfrak{S}_1$ stands for a set of entangled states.

Let ${\rm T} : \mathfrak{L}(\mathcal{H}) \rightarrow \mathfrak{L}(\mathcal{H})$ denotes transposition with respect to a fixed orthonormal basis in $\mathcal{H}$. Transposition is an example of a linear positive map (see Section \ref{MAPS}), that is, ${\rm T}(\mathfrak{L}_+(\mathcal{H})) \subset \mathfrak{L}_+(\mathcal{H})$. Moreover, it is trace-preserving which implies  ${\rm T}(\mathfrak{S}(\mathcal{H})) \subset \mathfrak{S}(\mathcal{H})$, i.e. it maps states into states. Let ${\rm id}_A : \HA \ra \HA$ be an identity map (${\rm id}_A(X) =X$ for any $X \in \mathfrak{L}(\HA)$). Define the partial transposition
\begin{equation}\label{}
  {\rm id}_A \otimes {\rm T} :  \mathfrak{T}(\HAB) \ra  \mathfrak{T}(\HAB)\ ,
\end{equation}
as follows
\begin{equation}\label{}
  [{\rm id}_A \otimes {\rm T}](a \otimes b) = a \otimes b^{\rm T}\ ,
\end{equation}
where $b^{\rm T} = {\rm T}(b)$. One usually denotes $X^\Gamma =  [{\rm id}_A \otimes {\rm T}]X$.

\begin{Definition} A positive operator  $X \in  \mathfrak{L}_+(\mathcal{H}_{AB})$ is Positive Partial Transpose ({\rm PPT}) iff $X^\Gamma \geq 0$, i.e. $X^\Gamma \in \mathfrak{L}_+(\HAB)$.
\end{Definition}
Clearly, if $\rho \in \mathfrak{S}(\HAB)$ is PPT, then $\rho^\Gamma \in \mathfrak{S}(\HAB)$ which means that $\rho^\Gamma$ is a legitimate state in $\HAB$.
One has the following

\begin{Proposition} A vector $\psi \in \HAB$ is separable iff the corresponding positive rank-1 operator $\,|\psi\>\<\psi|$ is PPT.
\end{Proposition}
For mixed states one has the celebrated

\begin{Theorem}[\cite{EW1,Peres}]\label{P-H} If a state $\rho \in \mathfrak{S}(\HAB)$ is separable, then $\rho$ is {\rm PPT}.
\end{Theorem}
Hence, an NPT (not PPT) state is necessarily entangled. The converse of the above Proposition not true, i.e. there exist PPT entangled states. However,

\begin{Theorem}[\cite{EW1}] If $D = d_A d_B \leq 6$,  then  $\rho \in \mathfrak{S}(\HAB)$ is separable if and only if  $\rho$ is {\rm PPT}.
\end{Theorem}
Actually, this result follows from much older result of St{\o}rmer \cite{Stormer-63} and Woronowicz \cite{Wor1} (cf. Proposition \ref{DEC-23}).
Now, let us introduce
\begin{equation}\label{}
  \mathfrak{L}^l := ({\rm id}_A \otimes {\rm T})\mathfrak{L}_l \ ,
\end{equation}
for $l=1,\ldots,d$, and define $\mathfrak{L}_k^l = \mathfrak{L}_k \cap \mathfrak{L}^l$, i.e. a convex cone a of positive PPT operators $X$  such that $\mathrm{SN}(X) \leq k$ and $\mathrm{SN}(X^\Gamma) \leq l$.   One finds
$\mathfrak{L}_{\rm SEP} = \mathfrak{L}_1 = \mathfrak{L}^1 = \mathfrak{L}_1^1$ and $\mathfrak{L}_d^d = \mathfrak{L}_{\rm PPT}$. Hence, the following hierarchy of inclusions follows
\begin{equation}\label{T-chain}
  \mathfrak{L}_{\rm SEP}  = \mathfrak{L}_1^1 \subset \mathfrak{L}_2^2 \subset\, \ldots\, \subset \mathfrak{L}_d^d = \mathfrak{L}_{\rm PPT} \subset \mathfrak{L}_+(\HAB)\ .
\end{equation}
Similarly, defining $\mathfrak{S}_k^l = \mathfrak{L}_k^l \cap \mathfrak{S}(\HAB)$ one has
\begin{equation}\label{}
    \mathfrak{S}_{\rm SEP} = \mathfrak{S}_1^1 \subset \mathfrak{S}_2^2  \subset \, \ldots\, \mathfrak{S}^d_d =  \mathfrak{S}_{\rm PPT}  \subset \mathfrak{S}(\HAB) \     .
\end{equation}
%Note, that $\mathfrak{S}_d \cap \mathfrak{S}^d $ is a convex set of PPT
%states in $\HAB$. Finally, $\mathfrak{S}_r \cap \mathfrak{S}^s$ is
%a convex subset of PPT states $\rho$  such that $\mathrm{SN}(\rho)
%\leq r$ and $\mathrm{SN}(\rho^\Gamma) \leq s$.
The hard problem in the entanglement theory is to check weather a given PPT state is entangled or separable. Note, that a set of PPT states is convex. It is no longer true for entangled PPT states: a convex combination of two entangled PPT states is clearly a PPT state but it needs not be entangled.
An important construction of PPT entangled states is based on  the following

\begin{Theorem}[Range criterion \cite{Range}] If $X$ is a separable operator in $\HAB$, then there exists a set of product vector $|a_i \> \otimes |b_i\> \in \HAB$ spanning the range of $X$ such that a set $|a_i \> \otimes |b_i^*\>$ spans the range of $X^\Gamma$.
\end{Theorem}
This theorem provides a powerful tool for constructing PPT entangled states.

\begin{Example}\label{EX-UPB} Let a set of normalized product vectors $|\alpha_i\> \otimes |\beta_i\> \in \mathbb{C}^d \otimes \mathbb{C}^d \, (i=1,\ldots,K < d^2)$ define so called unextendable product basis (UPB), i.e. there is no product vector in $\mathbb{C}^d \otimes \mathbb{C}^d$ which is orthogonal to all of them \cite{UPB1,UPB2}.
Define a projector
\begin{equation}\label{}
  \Pi_{\rm UPB} = \sum_{i=1}^K |\alpha_i\>\<\alpha_i| \ot
    |\beta_i\>\<\beta_i| \ ,
\end{equation}
that is, $\mathcal{H}_{\rm UPB} := \Pi (\mathbb{C}^d \otimes \mathbb{C}^d)$ is spanned by $|\alpha_i\> \otimes |\beta_i\>$. Define
\begin{equation}\label{XU1}
  X = \mathbb{I}_d \otimes \mathbb{I}_d - \Pi_{\rm UPB}\ ,
\end{equation}
which is the projection onto $\mathcal{H}_{\rm UPB}^\perp$.
It is evident that $X$ is {\rm PPT}. Indeed, one has
\begin{equation}\label{XU2}
  X^\Gamma = \mathbb{I}_d \otimes \mathbb{I}_d - \Pi_{\rm UPB}^\Gamma\ ,
\end{equation}
where $\Pi_{\rm UPB}^\Gamma = \sum_{i=1}^K |\alpha_i\>\<\alpha_i| \ot
    |\beta_i^*\>\<\beta_i^*|$ defines an orthogonal projector onto a subspace spanned by another UPB $|\alpha_i\> \otimes |\beta_i^*\>$.
It is therefore clear that  there is no product vectors in the range of $X$ and $X^\Gamma$ and hence due to the Range criterion $X$ can not be separable.
\end{Example}
This example may be generalized as follows

\begin{Definition} A PPT operator $X \in \mathfrak{L}_{\rm PPT}$ is called an edge operator if for arbitrary $Y \in \mathfrak{L}_{\rm SEP}$ an operator $X-Y$  is not PPT, i.e. either $X - Y \ngeq 0$ or $(X-Y)^\Gamma \ngeq 0$.
\end{Definition}
Range criterion immediately implies that an edge operator is necessarily entangled. A density operator being an edge operator is called an edge state.
Edge states lay on the boundary between PPT and NPT  states, and hence, they may be considered as most entangled PPT states.

\begin{Theorem}[\cite{opt}] Any PPT entangled operator $X$ may be decomposed as follows
\begin{equation}\label{}
  X = X_{\rm EDGE} + X_{\rm SEP}\ ,
\end{equation}
where $X_{\rm EDGE}$ and $X_{\rm SEP}$ is an edge and separable operator, respectively.
\end{Theorem}
Note, that $X$ defined in (\ref{XU1}) is by construction an edge operator.  There is, however, no general method to construct edge operators.

\section{Entanglement witnesses: definitions and basic properties}  \label{S-EW}

A positive operator  $X \in \mathfrak{L}_+(\HAB)$ satisfies $\<\Psi|X|\Psi\> \geq 0$ for all $\Psi \in \HAB$. The following definition provides a generalization of positivity

\begin{Definition} A Hermitian operator $W \in \mathfrak{L}(\HAB)$ is block-positive iff
\begin{equation}\label{}
  \< \psi \otimes \phi |W|\psi \otimes \phi\>  \geq 0 \ ,
\end{equation}
for all product vectors $\psi \otimes \phi \in \HAB$.
\end{Definition}
Let $\mathbb{W}_1$ denotes a set of block-positive operators in $\HAB$. Note, that $\mathbb{W}_1$ defines a convex cone in $\mathfrak{L}(\HAB)$.  It is clear  that $W \in \mathbb{W}_1$ iff $\tr (W X) \geq 0$ for all $X \in \mathfrak{L}_1=\mathfrak{L}_{\rm SEP}$. Let $\{e_1,\ldots,e_{d_A}\}$ be an orthonormal basis in $\HA$. Any operator $W \in \mathfrak{L}(\HAB)$ may be represented in the following block form
\begin{equation}\label{}
  W = \sum_{i,j=1}^{d_A} E_{ij} \otimes W_{ij}\ ,
\end{equation}
where $E_{ij} := |e_i\>\<e_j|$ and $W_{ij} \in \mathfrak{L}(\HB)$, that is, given the basis $\{e_k\}$ an operator $W$ may be represented as a $d_A \times d_A$ matrix with matrix elements (blocks) being operators from $\mathfrak{L}(\HB)$. Now, if $W$ is block-positive one has
\begin{equation}\label{}
  \< e_i \otimes \phi|W|e_i \otimes \phi\> = \< \phi|W_{ii}|\phi\> \geq 0 \ ,
\end{equation}
and hence diagonal blocks $W_{ii}$ are positive operators. Clearly, if $W$ is  positive, then $W$ is block-positive as well.

\begin{Definition}[\cite{Terhal}] An operator $W \in \mathfrak{L}(\HAB)$ is an entanglement witness (EW) if and only if it is block-positive but not positive, i.e. $W \in \mathbb{W}_1 - \mathfrak{L}_+(\HAB)$.
\end{Definition}

\begin{Example}[Flip operator] Let $\mathbb{F} \in \mathfrak{L}(\mathcal{H} \otimes \mathcal{H})$ be so called flip (or swap) operator defined as follows
\begin{equation}\label{}
  \mathbb{F}\, \psi \otimes \phi = \phi \otimes \psi\ .
\end{equation}
One has
\begin{equation}\label{}
  \< \psi \otimes \phi|\mathbb{F}|\psi \otimes \phi\> = |\< \psi|\phi\>|^2 \ ,
\end{equation}
and hence $\mathbb{F}$ is block-positive. If $\{e_1,\ldots,e_{d}\}$ is an orthonormal basis in $\mathcal{H}$, then
\begin{equation}\label{}
  \mathbb{F} = \sum_{i,j=1}^d E_{ij} \otimes E_{ji} \ ,
\end{equation}
which proves, that $ \sum_{i,j=1}^d E_{ij} \otimes E_{ji}$ is basis-independent.  Using
\begin{equation}\label{}
  \mathbb{F} = \sum_{i=1}^d E_{ii} \otimes E_{ii} + \sum_{i\neq j=1}^d E_{ij} \otimes E_{ji} \ ,
\end{equation}
one easily finds that $\mathbb{F}\, [e_i \otimes e_j - e_j \otimes e_i] = - [e_i \otimes e_j - e_j \otimes e_i]$ which proves that $\mathbb{F}$ is not positive and hence it defines an EW.
\end{Example}
The importance of entanglement witnesses is based on the following theorem arising from the celebrated Banach separation theorem

\begin{Theorem}[\cite{EW1,EW2}] A state $\rho \in \mathfrak{S}(\HAB)$ is entangled iff there exists an entanglement witness $W \in \mathfrak{L}(\HAB)$ such that $\tr (W\rho) <0$.
\end{Theorem}

\begin{Example}[Werner state \cite{WERNER}] Consider a Werner state in $\mathcal{H} \otimes \mathcal{H}$ defined by
\begin{equation}\label{Wer}
  \rho = p\, Q_S + (1-p)\, Q_A\ ,
\end{equation}
where $p\in [0,1]$ and
\begin{eqnarray*}
% \nonumber to remove numbering (before each equation)
  Q_S = \frac{2}{d(d+1)} \, [ \mathbb{I}_d \otimes \mathbb{I}_d + \mathbb{F} ] \ , \ \ \
  Q_A = \frac{2}{d(d-1)} \, [ \mathbb{I}_d \otimes \mathbb{I}_d - \mathbb{F} ] \ .
\end{eqnarray*}
It is well known that (\ref{Wer}) is separable iff it is PPT, that is, $p \geq 1/2$. Now,
\begin{equation}\label{}
  \tr (\rho \mathbb{F}) = 2(2p-1)\ ,
\end{equation}
which shows that $\mathbb{F}$ detects all entangled Werner states.
\end{Example}
Let us define a space of $k$-block-positive operators
\begin{equation}\label{}
  \mathbb{W}_k = \{ W \in \mathfrak{L}(\HAB)\, |\, \< \Psi | W| \Psi\> \geq 0 \ ; \ {\rm SR}(\Psi) \leq k \}  \ .
\end{equation}
Again, if $W \in \mathbb{W}_k$ then $W^\dagger = W$. Moreover, $\mathbb{W}_k$ defines a convex cone in $\mathfrak{L}(\HAB$).
One has
\begin{equation}\label{}
  \mathfrak{L}_+(\HAB) = \mathbb{W}_d \subset \mathbb{W}_{d-1} \subset \, \ldots\, \subset \mathbb{W}_1 \ .
\end{equation}
Similarly, let
\begin{equation}\label{}
  \mathbb{W}^k = ({\rm id}_A \otimes {\rm T})\mathbb{W}_k\ ,
\end{equation}
and finally
\begin{equation}\label{}
  \mathbb{W}_k^l  = \{ A + B \, |\, A \in \mathbb{W}_k \ , B \in \mathbb{W}^l \}\ ,
\end{equation}
which leads to the following chain of inclusions of convex cones
\begin{equation}\label{B-chain}
   \mathbb{W}_d^d \subset \mathbb{W}_{d-1}^{d-1} \subset \, \ldots\, \subset \mathbb{W}_1^1 = \mathbb{W}^1 = \mathbb{W}_1 \ .
\end{equation}
The above chain is dual to a similar chain defined in (\ref{T-chain}). One proves the following
\begin{Proposition} $W \in \mathbb{W}_k^l$ if and only if
%\begin{equation}\label{}
 $\, \tr (W X) \geq 0$
%\end{equation}
for all $X \in \mathfrak{L}_k^l$.
\end{Proposition}

\begin{Definition} An entanglement witness $W \in \mathbb{W}_1 - \mathfrak{L}_+(\HAB)$ is called

\begin{itemize}

\item $k$-Schmidt witness if $W \in \mathbb{W}_{k-1} - \mathbb{W}_k$,

\item decomposable  if $W \in \mathbb{W}^d_d - \mathfrak{L}_+(\HAB)$ and indecomposable otherwise,

\item atomic if $W \in  \mathbb{W}_1 - \mathbb{W}^2_2$.

\end{itemize}

\end{Definition}
It is clear that $\rho \in \mathfrak{S}(\HAB)$ has a Schmidt number ${\rm SN}(\rho) \geq k$ iff there exists a $k$-Schmidt witness $W$ such that ${\rm Tr}(W\rho) <0$ \cite{Sanpera1,Sanpera2}. Moreover, $W$ is an decomposable EW iff ${\rm Tr}(WX) \geq 0$ for all PPT states $\rho \in \mathfrak{S}_{\rm PPT}$. Hence, if $W$ is an EW and ${\rm Tr}(W\rho)<0$ for $\rho \in \mathfrak{S}_{\rm PPT}$, then $W$ is an indecomposable and $\rho$ is an entangled PPT state. Clearly, any decomposable EW has the following form
\begin{equation}\label{DEC}
  W = A + B^\Gamma\ ,
\end{equation}
with $A$ and $B$ being positive operators in $\HAB$. If $W$ is atomic, then it cannot be represented via (\ref{DEC}) with $A,B \in \mathfrak{L}_2$. It means that if $W$ is atomic and ${\rm Tr}(W\rho)<0$ for some PPT state $\rho$, then $\rho$ is entangled and ${\rm SN}(\rho) = {\rm SN}(\rho^\Gamma)=2$. It shows that atomic entanglement witnesses can detect the weakest form of quantum entanglement.

\begin{Proposition}[\cite{Stormer-63,Wor1}]\label{DEC-23} If $D=d_Ad_B \leq 6$, then all entanglement witnesses are decomposable.
\end{Proposition}

%\begin{Example} The flip operator $\mathbb{F}$ represents a decomposable EW. Indeed, one has

Note, that a flip operator provides a decomposable EW. Indeed, one finds
\begin{equation}\label{}
  \mathbb{F} = \sum_{i,j=1}^d E_{ij} \otimes E_{ji} = P^\Gamma\ ,
\end{equation}
where $P= \sum_{i,j=1}^d E_{ij} \otimes E_{ij} = d\,P^+_d$.  Due to (\ref{DEC}) the structure of decomposable entanglement witnesses is fully understood. Note, that decomposable entanglement witnesses define a convex set. It is no longer true for indecomposable ones. Their construction is not straightforward. Moreover, a convex combination of two indecomposable witnesses clearly provides a block-positive operator but it needs not be indecomposable.
An important construction of indecomposable witnesses is provided by the following

\begin{Example} Consider a PPT entangled operator $X$ defined in (\ref{XU1}) in terms of unextendible product basis $|\alpha_i\> \otimes |\beta_i\>$.
Following \cite{UPB1,UPB2} we provide a canonical indecomposable entanglement witness detecting $X$. Let us define
\begin{equation}\label{}
    W = \Pi_{\rm UPB} - d\, \epsilon|\Psi\>\<\Psi| \ ,
\end{equation}
where $|\Psi\>$ is a normalized maximally entangled vector in $\mathbb{C}^d \otimes \mathbb{C}^d$ such that
$\<\Psi|X|\Psi\> > 0$. A parameter $\epsilon$ is defined by
\begin{equation}\label{}
    \epsilon = \inf_{|\phi_1\> \otimes |\phi_2\>} \, \sum_{i=1}^K |\<\alpha_i|\phi_1\>|^2
    \<\beta_i|\phi_2\>|^2\ ,
\end{equation}
where the infimum is taken over all normalized product vectors
$|\phi_1\> \otimes |\phi_2\>$. One immediately finds that $W$ is block-positive (it follows from the well known fact that  $|\<\Psi|a \otimes b\>|^2 < 1/d$). Moreover,
\begin{equation}\label{}
  \tr (WX) = - \epsilon d \<\Psi|X|\Psi\> < 0 \ ,
\end{equation}
which proves that $W$ is an indecomposable EW.
\end{Example}
Remarkably, the above construction may be generalized to perform full characterization of indecomposable entanglement witnesses.

\begin{Theorem}[\cite{Lew1}] Any indecomposable entanglement witness $W \in \mathfrak{L}(\HAB)$ can be represented as follows:
\begin{equation}\label{}
  W = P + Q^\Gamma - \epsilon \, \mathbb{I}_{d_A} \otimes \mathbb{I}_{d_B}\ ,
\end{equation}
with
\begin{equation*}\label{}
\epsilon = \inf_{|a\>\otimes |b\>} \< a \otimes b|P + Q^\Gamma|a\otimes b\> \ ,
\end{equation*}
where the infimum is taken over all normalized product vectors
$|a\>\otimes |b\> \in \HAB$. Moreover, $P$ and $Q$ are positive operators such that
\begin{equation}\label{}
  \tr (P X_{\rm EDGE}) = \tr ( Q^\Gamma X_{\rm EDGE}) = 0\ ,
\end{equation}
for some edge operator $X_{\rm EDGE}$.

\end{Theorem}
In particular  $P$ and $Q$ might be  projectors onto the kernel of $X_{\rm EDGE}$ and $X^\Gamma_{\rm EDGE}$, respectively.

\section{Positive maps vs. entanglement witnesses }  \label{MAPS}

%In this section we provide a basic review on positive linear maps.
A linear map
\begin{equation}\label{Phi}
  \Phi : \mathfrak{L}(\HA) \rightarrow \mathfrak{L}(\HB)\ ,
\end{equation}
is called

\begin{itemize}

\item Hermitian (or Hermiticity-preserving) if $[\Phi(A)]^\dagger = \Phi(A^\dagger)$ for all $A \in \mathfrak{L}(\HA)$,

\item trace-preserving if $\tr \Phi(A) = \tr A$ for all $A \in \mathfrak{L}(\HA)$,

\item positive if $\Phi(A) \geq 0$ for any $A \geq 0$.

\end{itemize}
Actually, a positive map is necessarily Hermitian. A trace-preserving positive map (\ref{Phi}) maps $\mathfrak{S}(\HA)$ into $\mathfrak{S}(\HB)$, i.e. it maps quantum states living in $\HA$ into quantum states living in $\HB$. Note, that $\Phi$ is positive iff for any $\psi_A \in \HA$ and $\psi_B \in \HB$ one has
\begin{equation}\label{}
  \< \psi_B | \Phi(|\psi_A\>\<\psi_A|)|\psi_B\> \geq 0 \ .
\end{equation}

\begin{Definition} $\Phi$ is $k$-positive if
\begin{equation}\label{}
  {\rm id}_k \otimes \Phi : M_k(\mathbb{C}) \otimes \mathfrak{L}(\HA) \rightarrow M_k(\mathbb{C}) \ot\mathfrak{L}(\HB)\ ,
\end{equation}
is positive. $\Phi$ is completely positive (CP) if it is $k$-positive for $k=1,2,\ldots$.
\end{Definition}
It was observed by Choi \cite{Choi-CP} that $\Phi$ is CP iff it is $d$-positive ($d = \min\{d_A,d_B\}$). Remarkably, it turns out \cite{Choi-CP}
that to prove complete positivity it is enough to check positivity of ${\rm id}_A \otimes \Phi$ on a single element!

\begin{Proposition} A linear map  $\Phi : \mathfrak{L}(\HA) \rightarrow \mathfrak{L}(\HB)$ is completely positive if and only if
\begin{equation}\label{}
  [{\rm id}_A \otimes \Phi] P^+_{AA} \geq 0 \ ,
\end{equation}
where $P^+_{AA}$ is a maximally entangled state in $\HA \otimes \HA$.
\end{Proposition}
This result immediately implies the celebrated Kraus-Choi representation of a completely positive map
\begin{equation}\label{}
  \Phi(A) = \sum_k V_k A V_k^\dagger\ ,
\end{equation}
where $V_k : \HA \rightarrow \HB$. If $\{e_1,\ldots,e_{d_A}\}$ denotes an orthonormal basis in $\HA$ and $\Phi : \mathfrak{L}(\HA) \rightarrow \mathfrak{L}(\HB)$ is a linear map then  one defines a Choi matrix
\begin{equation}\label{CJ-1}
  C_\Phi := \sum_{i,j=1}^{d_A} E_{ij} \otimes \Phi(E_{ij}) \in M_{d_A}(\mathbb{C}) \otimes \mathfrak{L}(\HB) \backsimeq \mathfrak{L}(\HAB)\  .
\end{equation}
This way one establishes a correspondence between linear maps from $\mathfrak{L}(\HA)$ to $\mathfrak{L}(\HB)$ and bipartite operators from $\mathfrak{L}(\HAB)$. It should be stressed that this correspondence depends upon the basis  $\{e_1,\ldots,e_{d_A}\}$. One easily finds the inverse relation: for $X \in \mathfrak{L}(\HAB)$ one defines a linear map $\Phi$ as follows: if $X = \sum_{i,j} E_{ij} \otimes X_{ij}$, then $\Phi(E_{ij}) := X_{ij}$. This recipe is usually presented by
\begin{equation}\label{CJ-2}
  \Phi(a) = \tr_A [ a^{\rm T} \otimes \mathbb{I}_B \, X] \ ,
\end{equation}
where the transposition is performed with respect to $\{e_1,\ldots,e_{d_A}\}$. The correspondence defined by (\ref{CJ-1}) and (\ref{CJ-2}) is called Choi-Jamio{\l}kowski isomorphism \cite{Choi-CP,Jam}.

\begin{Remark} Actually, one may replace maximally entangled state $P^+_{AA}$ by arbitrary $P = |\psi\>\<\psi|$ such that $\psi \in \HA \otimes \HA$ has maximal Schmidt rank and still obtain the 1-1 correspondence between maps and linear operators in $\mathfrak{L}(\HAB)$.
\end{Remark}

Denote by $\mathfrak{P}_k$ a convex cone of $k$-positive map. One has the following chain of inclusions
\begin{equation}\label{}
  \mathfrak{P}_{\rm CP} = \mathfrak{P}_d \subset \mathfrak{P}_{d-1} \subset \ldots \subset \mathfrak{P}_1 =
  \mathfrak{P}(\mathfrak{L}(\HA),\mathfrak{L}(\HB))\ ,
\end{equation}
where by $\mathfrak{P}(\mathfrak{L}(\HA),\mathfrak{L}(\HB))$ we denote a convex cone of positive maps from $\mathfrak{L}(\HA)$ into $\mathfrak{L}(\HB)$.
If ${\rm T_A}$ is a transposition map (with respect to a fixed basis in $\HA$) acting in $\mathfrak{L}(\HA)$, then one defines
\begin{equation}\label{}
  \mathfrak{P}^k := \{ \Phi : \mathfrak{L}(\HA) \rightarrow \mathfrak{L}(\HB) \ |\ \Phi \circ {\rm T}_A \in \mathfrak{P}_k \}  \ .
\end{equation}
It is clear that equivalently one may define
\begin{equation}\label{}
  \mathfrak{P}^k := \{ \Phi : \mathfrak{L}(\HA) \rightarrow \mathfrak{L}(\HB) \ |\  {\rm T}_B \circ \Phi  \in \mathfrak{P}_k \} \ ,
\end{equation}
where ${\rm T_B}$ is a transposition map (with respect to a fixed basis in $\HB$) acting in $\mathfrak{L}(\HB)$.
Elements from $\mathfrak{P}^k$ are called $k$-copositive maps and maps from $\mathfrak{P}^d$ are called completely copositive.  Finally, let us introduce
\begin{equation}\label{}
  \mathfrak{P}_k^l = \{ \Phi + \Phi' \ | \ \Phi \in \mathfrak{P}_k \ \wedge \ \Phi' \in \mathfrak{P}^l \} \ .
\end{equation}
One says that a positive map $\Phi$ is

\begin{itemize}

\item decomposable if $\Phi \in \mathfrak{P}^d_d$ and indecomposable otherwise,

%\item $\Phi$ is indecomposable if $\Phi \in \mathfrak{P}_1 - \mathfrak{P}^d_d$,

\item atomic if $\Phi \in \mathfrak{P}_1 - \mathfrak{P}^2_2$.
\end{itemize}
Remarkably, the Choi-Jamio{\l}kowski isomorphism establishes elegant correspondence between cones $\mathfrak{P}_k^l$ of maps and the cones $\mathfrak{L}_k^l$ of operators. One proves  \cite{Patricot,KAROL,Ranade,Skowronek2} the following
\begin{Proposition} Let $\mathfrak{L}(\HA) \rightarrow \mathfrak{L}(\HB)$ be a linear map. The following statements are equivalent

\begin{enumerate}

\item   $\Phi \in \mathfrak{P}^l_k$,

\item
%\begin{equation}\label{}
 $ [{\rm id}_A \otimes \Phi]X \geq 0$
%\end{equation}
for all $X \in \mathfrak{L}_k^l \subset \mathfrak{L}(\HA \otimes \HA)$,

\item the Choi matrix
%\begin{equation}\label{}
 $ C_\Phi = \sum_{i,j=1}^{d_A} E_{ij} \otimes \Phi(E_{ij}) \in \mathfrak{L}_k^l \subset \mathfrak{L}(\HAB) \ . $
%\end{equation}

\end{enumerate}

\end{Proposition}
In particular $\Phi$ is positive iff $\sum_{i,j} E_{ij} \otimes \Phi(E_{ij})$ is block-positive.
Note, that if $\Phi$ is trace-preserving then $\Phi \in \mathfrak{P}^l_k$ iff
\begin{equation}\label{}
  [{\rm id}_A \otimes \Phi](\mathfrak{S}_k^l) \subset \mathfrak{S}(\HAB) \ .
\end{equation}

\begin{Remark}\label{R-Pillis}

The isomorphism between the vector space $\mathcal{L}(\mathfrak{L}(\mathcal{H}_A),\mathfrak{L}(\mathcal{H}_B))$ of linear maps and the vector space $\mathfrak{L}(\HAB)$ of bipartite operators was originally established by de Pillis \cite{Pillis}.  Note, that    $\mathcal{L}(\mathfrak{L}(\mathcal{H}_A),\mathfrak{L}(\mathcal{H}_B))$ may be equipped with a natural inner product
\begin{equation}\label{}
  \<\< \Phi|\Psi\>\> := \sum_{\alpha=1}^{D_A} \< \Phi(E_\alpha)|\Psi(E_\alpha)\>_{\rm HS} = \sum_{\alpha=1}^{D_A} {\rm Tr}[ \Phi(E_\alpha)^\dagger \Psi(E_\alpha)] \ ,
\end{equation}
where  $\{E_\alpha\}$ ($\alpha=1,\ldots,D_A=d_A^2$) is an arbitrary orthonormal basis in $\mathfrak{B}(\mathcal{H}_A)$, i.e. $\<E_\alpha|E_\beta\>_{\rm HS} = \delta_{\alpha\beta}$. It is clear that the above formula does not depend upon $\{E_\alpha\}$. Now, following
\cite{Pillis} one introduces a linear map $\mathcal{J} : \mathcal{L}(\mathfrak{L}(\mathcal{H}_A),\mathfrak{L}(\mathcal{H}_B)) \rightarrow \mathfrak{L}(\HAB)$ defined by the following relation
\begin{equation}\label{}
  \< \mathcal{J}(\Phi)| a^\dagger \otimes b\>_{\rm HS} = \< \Phi(a)|b\>_{\rm HS} \ ,
\end{equation}
for any $a \in \mathfrak{L}(\mathcal{H}_A)$ and $b \in \mathfrak{L}(\mathcal{H}_B)$. One easily finds
\begin{equation}\label{J}
  \mathcal{J}(\Phi)  = \sum_{\alpha=1}^{D_A} E_\alpha^\dagger \otimes \Phi(E_\alpha)\ ,
\end{equation}
where $\{E_\alpha\}$ is an arbitrary orthonormal basis in $\mathfrak{L}(\mathcal{H}_A)$. It should be stressed that $\mathcal{J}(\Phi)$ does not depend upon $\{E_\alpha\}$. Indeed, if $\{F_\alpha\}$ is another orthonormal basis, then $E_\alpha = \sum_\beta U_{\alpha\beta} F_\beta$ with $U_{\alpha\beta}$ being $D_A \times D_A$ unitary matrix. One has
\begin{equation}\label{J}
\fl
  \mathcal{J}(\Phi)  = \sum_{\alpha=1}^{D_A} E_\alpha^\dagger \otimes \Phi(E_\alpha) = \sum_{\alpha,\beta,\gamma=1}^{D_A} \overline{U}_{\alpha\beta} U_{\alpha\gamma} \, E_\beta^\dagger \otimes \Phi(E_\gamma) = \sum_{\alpha=1}^{D_A} F_\alpha^\dagger \otimes \Phi(F_\alpha) \ ,
\end{equation}
where we have used $\sum_{\alpha=1}^{D_A} \overline{U}_{\alpha\beta} U_{\alpha\gamma}  = \delta_{\beta\gamma}$.  In particular if $E_\alpha := E_{ij} =|e_i\>\<e_j|$ with $\{e_1,\ldots,e_{d_A}\}$ being an arbitrary orthonormal basis in $\mathcal{H}_A$, then
\begin{equation}\label{J}
  \mathcal{J}(\Phi)  = \sum_{i,j=1}^{d_A} E_{ij} \otimes \Phi(E_{ji})\ .
\end{equation}
Note, that (\ref{J}) differs from (\ref{CJ-1}) only by a partial transposition $\mathcal{J}(\Phi) = [{\rm T}_A \otimes {\rm id}_B]C_\Phi$. This minor difference produces important mathematical implication.

\begin{Proposition}[\cite{Pillis}] The linear map $\mathcal{J}$ provides an isometric isomorphism of two Hilbert spaces $\mathcal{L}(\mathfrak{L}(\mathcal{H}_A),\mathfrak{L}(\mathcal{H}_B))$ and $\mathfrak{L}(\HAB)$, that is,
\begin{equation}\label{}
\<\< \Phi|\Psi\>\> =   \< \mathcal{J}(\Phi)|\mathcal{J}(\Psi)\>_{\rm HS}  \ .
\end{equation}
for any pair of maps $\Phi,\Psi$.
\end{Proposition}
Indeed, one has
\begin{equation*}\label{}
\fl
  \< \mathcal{J}(\Phi)|\mathcal{J}(\Psi)\>_{\rm HS}  = \sum_{\alpha,\beta=1}^{D_A} \< E_\alpha^\dagger \otimes \Phi(E_\alpha)|   E_\beta^\dagger \otimes \Psi(E_\beta)\>_{\rm HS} = \sum_{\alpha=1}^{D_A} \< \Phi(E_\alpha)| \Psi(E_\alpha)\>_{\rm HS}= \<\< \Phi|\Psi\>\>\ .
\end{equation*}
De Pillis showed \cite{Pillis} that $\Phi$ is Hermitian iff $\mathcal{J}(\Phi)$ is Hermitian. Moreover, it turns out \cite{Pillis} that if $\mathcal{J}(\Phi) \geq 0$, then $\Phi$ is a positive map. Remarkably, Jamio{\l}kowski observed \cite{Jam} that one may relax positivity of  $\mathcal{J}(\Phi)$ to block-positivity.
\end{Remark}
On the recent discussion about Choi-Jamio{\l}kowski isomorphism and the one introduced originally by de Pillis see the recent paper \cite{Luo}.
The analysis of the structure of positive maps may be extended to an infinite dimensional case \cite{Paulsen,Stormer-2013}. The infinite dimensional generalization of the Choi-Jamio{\l}kowski isomorphism was recently provided by Holevo \cite{Holevo}. For more detailed discussion about positive maps see \cite{Bhatia,Paulsen,Kye-REV,Stormer-63,Stormer-2013,Wor1,Wor2,Skowronek1,Skowronek2,Eom,DC1}.

\section{From separability criteria to entanglement witnesses}  \label{SEP-EW}

In this section we show how well known separability criteria enables one to construct an appropriate entanglement witnesses.

\subsection{Positive maps}

%It is clear that  if $\Phi : \mathfrak{L}(\HA) \rightarrow \mathfrak{L}(\HB)$ and $\Phi' : \mathfrak{L}(\HB) \rightarrow \mathfrak{L}(\mathcal{H}_C)$ are %positive maps, then $\Phi' \circ \Phi : \mathfrak{L}(\HA) \rightarrow \mathfrak{L}(\mathcal{H}_C)$ is positive as well.
Note that if $\Phi_1$ and $\Phi_2$ are two positive maps from $\mathfrak{L}(\mathcal{H}_{A_i})$ to  $\mathfrak{L}(\mathcal{H}_{B_i})$ $(i=1,2$), then
\begin{equation}\label{}
  \Phi_1 \otimes \Phi_2 : \mathfrak{L}(\mathcal{H}_{A_1} \otimes \mathcal{H}_{A_2}) \rightarrow \mathfrak{L}(\mathcal{H}_{B_1} \otimes \mathcal{H}_{B_2})\ ,
\end{equation}
needs not be positive (clearly $\Phi_1 \otimes \Phi_2$ is positive if both $\Phi_1$ and $\Phi_2$ are CP). This simple mathematical observation has profound physical significance.
The importance of positive maps in entanglement theory is based on  the following

\begin{Theorem}[\cite{EW1}] A positive operator $X \in \mathfrak{L}_+(\HAB)$ is separable if and only if
\begin{equation}\label{}
  [{\rm id}_A \otimes \Lambda]X \geq 0 \ ,
\end{equation}
for all positive maps $\Lambda : \mathfrak{L}(\HB) \rightarrow \mathfrak{L}(\HA)$.
\end{Theorem}
Indeed, if $X$ is separable, i.e. $X = \sum_k A_k \otimes B_k$ with positive $A_k \in \mathfrak{L}_+(\HA)$ and $B_k \in \mathfrak{L}_+(\HB)$, then  $[{\rm id}_A \otimes \Lambda]X = \sum_k A_k \otimes \Lambda(B_k) \geq 0\,$ for any positive map $\Lambda$. Conversely, if $[{\rm id}_A \otimes \Lambda]X \geq 0$, then $\tr ( P^+_{AA} [{\rm id}_A \otimes \Lambda]X ) \geq 0$. Now, denote by $\Lambda^\#$ a dual map
\begin{equation}\label{}
  \Lambda^\# : \mathfrak{L}(\HA) \rightarrow \mathfrak{L}(\HB)\ ,
\end{equation}
defined by
\begin{equation}\label{}
  \tr [ \Lambda^\#(A) \cdot B] := \tr [A \cdot\Lambda(B)] \ .
\end{equation}
Note, that $\Lambda^\#$ is positive iff $\Lambda$ is positive. Now, due to the Choi-Jamio{\l}kowski isomorphism $W = [{\rm id}_A \otimes \Lambda^\#]P^+_{AA}$ is block-positive and hence $\tr (W X) \geq 0$ for all block-positive operators $W$ which implies that $X$ is separable.

\begin{Remark}
Actually, it is enough to consider only trace-preserving positive maps.  Hence, a state $\rho \in \mathfrak{S}(\HAB)$ is separable iff
\begin{equation}\label{}
  [{\rm id}_A \otimes \Lambda]\rho \in \mathfrak{S}(\mathcal{H}_{A}\otimes \HA) \ ,
\end{equation}
for all trace-preserving positive maps $\Lambda : \mathfrak{S}(\HB) \rightarrow \mathfrak{S}(\HA)$.
\end{Remark}
For any positive (but not completely positive) map $\Lambda : \mathfrak{L}(\HA) \rightarrow \mathfrak{L}(\HB)$ one constructs the corresponding entanglement witness $W_\Lambda = [{\rm id}_A \otimes \Lambda]P^+_{AA} \in \mathfrak{L}(\HAB)$. Note, however, that even if $[{\rm id} \otimes \Lambda^\#]X \ngeq 0$ one may have $\tr (W_\Lambda X) \geq 0$, i.e. in general a positive map $\Lambda$ detects more entangled operators that the corresponding witness $W_\Lambda$.

\begin{Example}

Consider the transposition map ${\rm T}$ and the corresponding witness
$\mathbb{F}$. It is well known \cite{HHHH} that an isotropic state
\begin{equation}\label{}
 \rho_p =  \frac{p}{d^2} \mathbb{I}_d \otimes \mathbb{I}_d + (1-p)
\proj{\Psi^+_d}\ ,
\end{equation}
is separable iff it is PPT which is equivalent to $p\geq \frac d{d+1}$. Hence ${\rm T}$ detects all entangled isotropic states. On the other
hand, the mean value of corresponding witness $\mathbb{F}$ is equal
$\mathrm{Tr}(\rho \mathbb{F}) = \frac{p}{d^2} \mathrm{Tr} \mathbb{F} +
(1-p)\<\Psi_+ | \mathbb{F} \Psi_+\>  = \frac{p}{d} + (1-p) > 0$ and no isotropic state is
detected by $\mathbb{F}$.
\end{Example}
There is, however, a canonical way to assign an entanglement witness that detects $X \in \mathfrak{L}(\HAB)$ as well. Since $[{\rm id}_A \otimes \Lambda]X \ngeq 0$ this operator has at least one negative eigenvalue
\begin{equation}\label{}
  ([{\rm id}_A \otimes \Lambda]X) |\psi\> = \lambda |\psi\> \ ,
\end{equation}
with $\lambda < 0$ and $\psi \in \HA \otimes \HA$ is an entangled vector.  Observe that $W = [{\rm id}_A \otimes \Lambda^\#]|\psi\>\<\psi|$ is an entanglement witness such that $\tr (W X) < 0$. Indeed, one finds
\begin{equation*}\label{}
\fl
  \tr (W X) = \tr ( [{\rm id}_A \otimes \Lambda^\#]|\psi\>\<\psi| \cdot X ) = \tr ( |\psi\>\<\psi| \cdot [{\rm id}_A \otimes \Lambda]X ) = \lambda ||\psi||^2 < 0 \
\end{equation*}

\subsection{Realignment criterion}

Let us recall well known computable cross norm or realignment criterion \cite{RE2,RE1}. Since $\mathfrak{L}(\HAB) = \mathfrak{L}(\HA) \otimes \mathfrak{L}(\HB)$ is a tensor product of two Hilbert spaces we may apply Schmidt decomposition to any bipartite operator from $\mathfrak{L}(\HAB)$. In particular if $\rho$ is a density operator in $\HAB$, then
\begin{equation}\label{GG}
  \rho = \sum_k \lambda_k\, G^A_k \otimes G^B_k \ ,
\end{equation}
with $\lambda_k \geq 0$, and $G^A_k$ and $G^B_k$ are orthonormal basis in $\mathfrak{L}(\HA)$ and $\mathfrak{L}(\HB)$, respectively.

\begin{Theorem}[\cite{RE2,RE1}] If $\rho$ is separable, then
\begin{equation}\label{}
  \sum_k \lambda_k \leq 1\ .
\end{equation}
\end{Theorem}
Hence, if $\sum_k \lambda_k > 1$, then $\rho$ is necessarily entangled. Suppose now that we recognized the entanglement of $\rho$ checking that $\sum_k \lambda_k > 1$. The question is how to construct the corresponding entanglement witness $W$ such that $\tr (W \rho) < 0$. One easily proves

\begin{Proposition} If $\rho$ is Hermitian, then the operator Schmidt decomposition (\ref{GG}) implies that $G^A_k$ and $G^B_k$ are Hermitian as well.
\end{Proposition}
The corresponding witness is constructed as follows \cite{Guhne}
\begin{equation}\label{}
  W = \mathbb{I}_A \otimes \mathbb{I}_B - \sum_k  G^A_k \otimes G^B_k \ .
\end{equation}
One has
\begin{equation*}\label{}
  \< \alpha \otimes \beta|W|\alpha\otimes \beta\> = 1 - \sum_k  \< \alpha|G^A_k|\alpha\> \<\beta| G^B_k|\beta\>  \ ,
\end{equation*}
where we assumed normalization $||\alpha||=||\beta||=1$. Now, let
\begin{equation*}\label{}
  |\alpha\>\<\alpha|= \sum_{k} a_k G^A_k \ , \ \ \  |\beta\>\<\beta|= \sum_{l} b_l\, G^B_l \ ,
\end{equation*}
with $a_k,b_k \in \mathbb{R}$. Hence, $ \< \alpha \otimes \beta|W|\alpha\otimes \beta\> = 1 - \sum_k a_k b_k $.
Now, since $\sum_k a_k^2 = \sum_l b_l^2 = 1$ the Cauchy-Schwarz inequality implies $\sum_k a_k b_k \leq 1$.   Hence, $ \< \alpha \otimes \beta|W|\alpha\otimes \beta\> \geq 0$ which proves that $W$ is block-positive. Now
\begin{equation}\label{}
  \tr (W \rho) = 1 - \sum_k \lambda_k < 0\ ,
\end{equation}
which proves that $W$ is an entanglement witness detecting $\rho$.

\subsection{Bell inequalities and entanglement witnesses}

%Let us briefly discuss the relationship between Bell inequality derived by Clauser, Horne, Shimony, and Holt (CHSH) and  entanglement witnesses.
The original Bell inequality, which is based on the perfect anti-correlations of the so-called singlet state, was
 extended by Clauser, Horne, Shimony, and Holt (CHSH) \cite{CHSH} to a more general inequality for two observers
each having the choice of two measurement settings with two outcomes. The violation of a Bell inequality implies the nonexistence
of a local hidden variable (LHV) model for the correlations observed with respect to a certain quantum state. The standard setting for the CHSH inequality reads as follows: a source emits two particles -- one particle to each of two receivers -- and each receiver can perform one out of two dichotomic measurements $A_i$ and $B_i$ ($i=1,2$), respectively. Now, if the physical process can be described by a LHV model, the following inequality
\begin{equation}\label{CH}
  \mathbb{E}(A_1,B_1) +  \mathbb{E}(A_1,B_2) +  \mathbb{E}(A_2,B_1) -  \mathbb{E}(A_2,B_2) \leq 2\ ,
\end{equation}
derived by CHSH has to be satisfied, where $\mathbb{E}(A_i,B_j)$ denotes  the expectation
value of the correlation experiment $A_i \otimes B_j$.

For the two qubit case one introduces the following CHSH operator
\begin{equation}\label{}
  B_{\rm CHSH} = \mathbf{a}_1\mbox{\boldmath $\sigma$}  \otimes (\mathbf{b}_1 + \mathbf{b}_2)\mbox{\boldmath $\sigma$} + \mathbf{a}_2\mbox{\boldmath $\sigma$} \otimes (\mathbf{b}_1 - \mathbf{b}_2)\mbox{\boldmath $\sigma$}\ ,
\end{equation}
where $\mathbf{a}_1,\mathbf{a}_2,\mathbf{b}_1,\mathbf{b}_2 \in \mathbb{R}^3$ and $\mbox{\boldmath $\sigma$} = (\sigma_1,\sigma_2,\sigma_3)$ is the vector of Pauli matrices.  The CHSH inequality requires that
\begin{equation}\label{}
  \tr (B_{\rm CHSH} \cdot \rho_{\rm LHV}) \leq 2 \ ,
\end{equation}
is fulfilled for all  two qubit states $\,\rho_{\rm LHV}$ admitting an LHV model. One defines as a CHSH witness the witness which is positive
on all LHV states and which can be constructed as follows
\begin{equation}\label{}
  W_{\rm CHSH} = 2 \mathbb{I}_2 \otimes \mathbb{I}_2 - B_{\rm CHSH}\ .
\end{equation}
Strictly speaking $W_{\rm CHSH}$ is a ``non-locality witness'', that is, if $\tr (W_{\rm CHSH} \cdot \rho) < 0$ then $\rho$ does not admit an LHV model. Any such state is necessarily entangled and hence $W_{\rm CHSH}$ is a legitimate entanglement witness.

Note, that any bipartite Bell inequality may be represented as follows
\begin{equation}\label{}
  \sum_{i,j} \lambda_{ij}\, \mathbb{E}(A_i,B_j) \leq c \ ,
\end{equation}
for two families of local observables $A_i \in \mathfrak{L}(\HA)$ and $B_j \in \mathfrak{L}(\HB)$. Therefore, the corresponding witness may be constructed as follows
\begin{equation}\label{}
  W = c\, \mathbb{I}_A \otimes \mathbb{I}_B - \sum_{i,j} \lambda_{ij}\, A_i \otimes B_j\ .
\end{equation}

\begin{Remark} It was conjectured by Peres that for a bipartite  scenario a PPT state can not violate any Bell inequality. If the conjecture is true then all entangled witnesses constructed out of the Bell inequalities  are decomposable.
\end{Remark}
For further analysis see \cite{Hyllus-Bell,EW2}.

\subsection{Entanglement witnesses from mutually unbiased bases }

Two ortonormal bases $\{|e_i\> \},\{ |f_j\> \}$ in $\mathbb{C}^d$ are called \textit{mutually unbiased} (MUB) \cite{schwinger} iff for all $i,j=1,\ldots,d$ one has $ |\langle e_i | f_j \rangle|^2 = \frac 1d$. We call a set of orthonormal basis $\{\mathcal{B}_1,\ldots,\mathcal{B}_m\}$ mutually unbiased, if any two of them $\mathcal{B}_i$ and $\mathcal{B}_j$  are mutually unbiased. There are at most $d+1$ mutually unbiased bases in $\mathbb{C}^d$ \cite{wooters} and if $d$ is a power of prime then this bound is saturated. For $d=2$ one finds three mutually unbiased bases
\begin{equation}\label{MUB-2}
\fl
  \mathcal{B}_1 = \{ e_0,e_1\}\ , \ \ \mathcal{B}_2 = \left\{ \frac{e_0 + e_1}{\sqrt{2}}, \frac{e_0 - e_1}{\sqrt{2}}\right\}\ , \ \
  \mathcal{B}_3 =\left\{ \frac{e_0 + ie_1}{\sqrt{2}}, \frac{e_0 - ie_1}{\sqrt{2}}\right\}\ ,
\end{equation}
which are eigenbasis of Pauli matrices $\sigma_3$, $\sigma_1$ and $\sigma_2$, respectively.
Now, for any two sets of MUBs $\{\mathcal{B}_1,\ldots,\mathcal{B}_m\}$ and $\{\widetilde{\mathcal{B}}_1,\ldots,\widetilde{\mathcal{B}}_m\}$  in $\mathbb{C}^d$ one defines \cite{Hiesmayr}
\begin{equation}\label{}
  \mathcal{I}_m(\rho) := \sum_{\alpha=1}^m \sum_{i=1}^d \< e^{(\alpha)}_i \otimes f^{(\alpha)}_i | \rho| e^{(\alpha)}_i \otimes f^{(\alpha)}_i\> \ ,
\end{equation}
where $e^{(\alpha)}_i \in \mathcal{B}_\alpha$ and $f^{(\alpha)}_j \in \widetilde{\mathcal{B}}_\alpha$.

\begin{Proposition}[\cite{Hiesmayr}] If $\rho$ is separable, then
\begin{equation}\label{}
\mathcal{I}_m(\rho) \leq 1 + \frac{m-1}{d} \ .
\end{equation}
for $m \leq d+1$.
\end{Proposition}
It is, therefore, clear that the following bipartite operator
\begin{equation}\label{}
  W_m :=  \lambda_m \, \mathbb{I}_d \otimes \mathbb{I}_d -  \sum_{\alpha=1}^m \sum_{i=1}^d |e^{(\alpha)}_i\>\<e^{(\alpha)}_i|  \otimes |f^{(\alpha)}_i \>   \<f^{(\alpha)}_i| \ ,
\end{equation}
with $\lambda_m = 1 + \frac{m-1}{d}$ is block-positive. Hence, if $W \ngeq 0$, then it provides a legitimate entanglement witness.

\begin{Example} For $d=2$ let us take $\{\mathcal{B}_1,\mathcal{B}_2,\mathcal{B}_3\}$ defined in (\ref{MUB-2}) and $\widetilde{\mathcal{B}}_k = \mathcal{B}_k^*$.  One easily finds for $m=3$
\begin{equation}\label{}
  W_3 = \mathbb{I}_2 \otimes \mathbb{I}_2 - 2P^+_2\ .
 % \left( \begin{array}{cc|cc} . & . & . & -1 \\ . & 1 & . & . \\ \hline . & . & 1 & . \\ -1 & . & . & . \end{array} \right)\ .
\end{equation}
Interestingly, $W_3$ detects all entangled two qubit isotropic states.
\end{Example}
One shows \cite{Hiesmayr} that already for $m=2$ all pure entangled states are detected and if $m=d+1$ all entangled isotropic states in $\mathbb{C}^d \otimes \mathbb{C}^d$ are detected as well.

%\begin{Remark}

\subsection{Entanglement witnesses form the convex optimization problem}

Very efficient  method for detecting entanglement is provided by the complete family of separability criteria
introduced by Doherty et. al. \cite{Doherty-1,Doherty-2}. Basically, this method relies on a hierarchical characterisation of separable
states: any bi-partite separable state allows for PPT symmetric $n$-partite extension with arbitrary $n >2$ (see \cite{Doherty-1,Doherty-2} for details). This method has a number of very appealing features:

\begin{enumerate}

\item The set of criteria is complete, i.e. all entangled states are detected at some stage (when the extension from $n$ to $(n+1)$ parties is no longer possible).

\item  The criteria can be cast into a semidefinite program which is a convex optimisation problem for which efficient algorithms
exist.

\item  Remarkably, when a state is found entangled, this algorithm automatically yields an entanglement witness for that
state. These entanglement witnesses turn out to be of a special form, namely the bi-hermitian form
\begin{equation}\label{}
  \< x \otimes y|W|x \otimes y\> = \sum_{k,l=1}^{d_A} \sum_{\mu,\nu=1}^{d_B}\, W_{k\mu;l\nu}\, x^*_k\, x_l\, y^*_\mu\, y_\nu \ ,
\end{equation}
can be written as sums of squares (cf. \cite{Terhal}) and hence the block-positivity immediately follows.
\end{enumerate}

%\end{Remark}

\section{Optimality, extremality and exposedness}  \label{S-OPTIMAL}

\subsection{Optimal entanglement witnesses}

Given an entanglement witness $W$ one defines a set of all entangled states in $\HAB$ detected by $W$
\begin{equation}\label{}
  D_{W} = \{ \, \rho \in \mathfrak{S}(\HAB)\, |\, {\rm Tr}(\rho W) < 0 \, \}\ .
\end{equation}
Suppose now that we are given two entanglement witnesses $W_1$ and $W_2$ in $\HAB$.

\begin{figure}[!h] \label{FIG-OPT}
%  \begin{minipage}[t]{5cm}
%  \vspace{0pt}
\begin{center}
 \includegraphics[width=7cm]{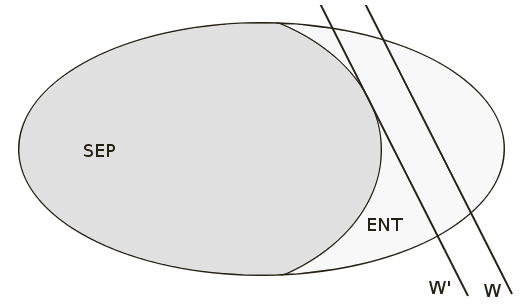}
\end{center}
\caption{Two entanglement  witnesses: not optimal $W$ and optimal $W'$.
 }
 \end{figure}

\begin{Definition}[\cite{opt}] \label{D-OPT}
We call $W_1$ finer than $W_2$ if $D_{W_1} \supseteq  D_{W_2}$. $W$ is called optimal if there is no other entanglement witness which is finer than $W$.
\end{Definition}
One proves \cite{opt} the following
\begin{Proposition}
$W$ is optimal EW iff for any  positive operator $P$ an operator $W - P$ is no longer block-positive.
\end{Proposition}
Roughly speaking an optimal witness cannot be `improved' (cf. Fig. \ref{FIG-OPT}). It is, therefore, clear that optimal witness are sufficient to detect all entangled states.

\begin{Remark} Let us observe that if $P$ is a positive operator $P \in \mathfrak{L}_+(\HAB)$, then
\begin{equation}\label{}
  W_\lambda = P - \lambda\, \mathbb{I}_A \otimes \mathbb{I}_B\ ,
\end{equation}
is block-positive whenever
\begin{equation}\label{ll0}
  \lambda \leq \lambda_0 = \inf_{|\alpha\>\ot|\beta\>} \< \alpha\otimes \beta|P|\alpha\otimes \beta\>\ ,
\end{equation}
where the infimum is performed over normalized vectors $|\alpha\>$ and $|\beta\>$. Hence, if $W_\lambda$ is not a positive operator and $\lambda \leq \lambda_0$, then $W_\lambda$ is a legitimate entanglement witness. Some authors claim that $W_{\lambda_0}$ is optimal. In general is is not true. Note, that $W_{\lambda_0}$ belongs to the boundary of the set of block-positive operators, however, it needs not be optimal in the sense of Definition \ref{D-OPT}.
\end{Remark}

Authors of \cite{opt} provided the following sufficient condition of optimality: for a given entanglement witness $W$ define
\begin{equation*}\label{SPAN}
  P_W = \{\, |\psi \otimes \phi\> \in \mathcal{H}_A \otimes \mathcal{H}_B\, |\, \< \psi \otimes \phi |W|\psi \otimes \phi\>  = 0 \, \} \ .
\end{equation*}
We say that $W$ possesses a {\em spanning property}  if ${\rm span} P_W = \HAB$.

\begin{Proposition}[\cite{opt}]
Any entanglement witness possessing a spanning property is optimal.
\end{Proposition}

\begin{Example} Flip operator possesses spanning property. One has
\begin{equation}\label{}
  \< \psi \otimes \phi|\mathbb{F}|\psi \otimes \phi\> =  \< \psi \otimes \phi|\phi \otimes \psi\> = |\<\psi|\phi\>|^2\ .
\end{equation}
One checks that the following set of product vectors:
$$  e_k \otimes e_l  \ \  (k \neq l)\ ,\ \ \   (e_m + ie_n)\otimes (e_m - ie_n)\ \ (m < n) \ ,  $$
span $\mathbb{C}^d \otimes \mathbb{C}^d$. Hence, flip operator provides an optimal EW.
\end{Example}

If $P_W$ does not span $\HA \otimes \HB$, then for a vector
$\phi\in {\rm span} P_W^\perp$ one can consider a witness $W' = W - \lambda
|\phi\>\<\phi|$, where
\begin{equation}\label{}
  \lambda = \inf_{|a\> \otimes |b\>} \frac{\< a \otimes b
| W| a \otimes b \>}{|\< a \otimes b| \phi \>|^2}\ .
\end{equation}
 If $\lambda>0$ then the
product vector $|a\>\otimes |b\>$  which realizes the above infimum belongs to $P_{W'}$
and hence $P_{W'}$ spans the bigger space than $P_W$. Repeating this procedure
one can construct a witness with a spanning property and hence optimal. If $\lambda=0$ for all $\phi \in {\rm span} P_W^\perp$, then it is
impossible to subtract a positive observable from the witness and it is
optimal but without a spanning property. Examples of such witnesses will be presented in section \ref{S-DIAG}.

\begin{Definition}
A linear subspace $S$ in $\HA \otimes \HB$ is called completely entangled (CES) if there is no product vector in $S$.
\end{Definition}
It turn out that an optimal decomposable entanglement witness $W$ has the form $W = Q^\Gamma$, where $Q$ is a positive operator supported on some {\rm CES} in $\HA \otimes \HB$.

\begin{Proposition}[\cite{Tura}]\label{Tura}
For a decomposable entanglement witness $W$ acting
in $\HA \otimes \HB$ such that $d=\min\{d_A,d_B\} =2$ the following three conditions
are equivalent:
\begin{enumerate}
 \item $W$ has a  spanning property,
 \item $W$ is optimal,
 \item $W = Q^\Gamma$, where $Q$ is a positive operator supported on some {\rm CES} in $\HA \otimes \HB$.
\end{enumerate}
\end{Proposition}
If $d > 2$, then
then $(iii)$ does not imply $(i)$ (cf. \cite{Tura}), $(ii)$ does not imply $(i)$ (cf. \cite{AugGS}), and $(iii)$ does not imply $(ii)$ (cf. \cite{KyeCES}).

Consider now an indecomposable witness $W$ and define
\begin{equation}\label{}
  D^{\rm PPT}_{W} = \{ \, \rho \in \mathfrak{S}_{\rm PPT} \,  |\, {\rm Tr}(\rho W) < 0 \, \}\ ,
\end{equation}
i.e. a set of PPT entangled states detected by $W$.

\begin{Definition}[\cite{opt}] \label{D-nd-OPT}
We call $W_1$ nd-finer than $W_2$ if $D^{\rm PPT}_{W_1} \supseteq  D^{\rm PPT}_{W_2}$. $W$ is called nd-optimal if there is no other entanglement witness which is nd-finer than $W$.
\end{Definition}
One proves \cite{opt} the following
\begin{Proposition}
$W$ is nd-optimal EW iff for any  decomposable operator $D = A + B^\Gamma$ ($A,B\geq 0$) an operator $W - D$ is no longer block-positive.
\end{Proposition}
It is clear that nd-optimality is stronger than optimality.

\begin{Proposition}[\cite{opt}]
An entanglement witness $W$ is nd-optimal if and only if both $W$ and $W^\Gamma$ are optimal.
\end{Proposition}

\begin{Remark} Recently Ha and Kye \cite{cospan}  proposed the following terminology:

\begin{itemize}

\item $W$ has a co-spanning property iff $W^\Gamma$ has a spanning property,

\item $W$ is co-optimal iff $W^\Gamma$ is optimal,

\item $W$ is bi-optimal iff $W$ is optimal and co-optimal.

\end{itemize}
Hence, nd-optimality is equivalent to bi-optimality.

\end{Remark}
Note, that if $W$ is optimal and decomposable, then $W^\Gamma \geq 0$ and hence it is no longer a witness. Now, if $W$ is indecomposable and optimal, then $W^\Gamma$ is indecomposable as well but needs not be optimal. In this case $W$ is optimal but not nd-optimal. An example of optimal but not nd-optimal witness was provided in \cite{cospan}

%%%%%%%%%%%%%%%%%%%%%%% SPA %%%%%%%%%%%%%
\begin{Remark} It is well known that a positive but not CP map can not be realized as a physical operation. One defines a Structural Physical Approximations (SPA) to a positive map $\Phi$ as follows
\begin{equation}\label{}
  \Phi_p(A) := p\, \Phi(A) + (1-p) \mathbb{I}_B \tr A\ ,
\end{equation}
where $p$ is such that $\Phi_p$ is CP. Few years ago authors of \cite{SPA} posed the following conjecture: SPA to optimal positive maps are entanglement breaking channels. In the language of entanglement witnesses this conjecture may be reformulated as follows: Let $W$ be a normalized (i.e. $\tr W=1$) entanglement witness  acting on a Hilbert space $\mathcal{H}=\mathcal{H}_A \otimes \mathcal{H}_B$ of finite dimension $D=d_A  d_B$. Consider the following convex combination
\begin{equation}\label{}
  W(p)  = pW + \frac{1-p}{D}\, \mathbb{I}_A \otimes \mathbb{I}_B\ ,
\end{equation}
i.e. a mixture of  $W$ with a maximally mixed state.  It is clear that there exists the largest $p_* \in (0,1)$ such that for all $p \leq p_*$ the operator $W(p) \geq 0$ and hence it defines a legitimate quantum states. One calls $W(p_*)$ the Structural Physical Approximations of $W$. SPA conjecture \cite{SPA} states that if $W$ is an optimal EW, then its SPA $W(p_*)$ defines a separable state. It is clear that if $W(p_*)$ is separable, then for all $p\leq p_*$ a state $W(p)$ is separable as well. This conjecture was supported by numerous examples of witnesses (see \cite{DC7,DC9,DC10,Justyna-2013,Zwolak-rec,Qi2} and Sections \ref{S-DIAG}, \ref{S-OPT}) and also analyzed in the continuous variables case \cite{A}. Recently SPA conjecture has been disproved by Ha and Kye \cite{Ha-JMP} for indecomposable witnesses and in \cite{DC-SPA} for decomposable ones (see also recent discussion in \cite{D1,D2,Ha-Yu,Ha-2012,Qi2,Chiny}).

\end{Remark}

\subsection{Extremal entanglement witnesses}

Two block-positive operators $W_1$ and $W_2$ are equivalent $(W_1 \backsim W_2$) iff $W_1 = a W_2$ for some $a > 0$. It is clear that if $W_1$ and $W_2$ are entanglement witnesses and $W_1 \backsim W_2$, then $D_{W_1} = D_{W_2}$, i.e. $W_1$ and $W_2$  have the same ``detection power''. Geometrically it means that $W_1$ and $W_2$ belong to the same ray in the convex cone of block-positive operators $\mathfrak{L}_1(\HAB)$.

\begin{Definition} A block-positive operator  $W$ is extremal if $W - W'$ is no longer block-positive, where $W'$ is an arbitrary block-positive operator not equivalent to $W$.
\end{Definition}
To avoid rays one may prefer to work with normalized operators $\tr W=1$. Normalized block-positive operators define a compact convex set in $\mathbb{R}^{D^2-1}$ ($D=d_Ad_B$). Then $W$ is extremal if it is an extremal point of this compact convex set. By the celebrated Krein-Milman theorem \cite{conv} a compact convex set is uniquely defined in terms of its extremal points. Note, that any extremal entanglement witness is optimal.

\begin{Proposition} If $W$ is an extremal decomposable entanglement witness if and only if $W = |\psi\>\<\psi|^\Gamma$ and $|\psi\>$ is an entangled vector in $\HAB$.
\end{Proposition}

\begin{Example} A flip operator $\mathbb{F}$ in $\mathbb{C}^d \otimes \mathbb{C}^d$ satisfies $\mathbb{F} = d P^{+\Gamma}_d$ and hence $\mathbb{F}$ provides an extremal entanglement witness.
\end{Example}

\begin{Proposition} $W$ be an indecomposable extremal entanglement witness,  then $W^\Gamma$ is also extremal and indecomposable.
\end{Proposition}
Hence, any indecomposable extremal entanglement witness is nd-optimal. For examples of indecomposable extremal entanglement witness see section \ref{S-DIAG}.
Among extremal block-positive operators there is a class of so called exposed operators.

\begin{Definition}
An extremal block-positive operator $W_0 \in \mathfrak{L}_1$ is exposed if and only if it enjoys the following property: if there exists a separable operator $X_{\rm sep}$ such that $\tr (W_0 X_{\rm sep})=0$ and $\tr (W X_{\rm sep})=0$ for some block-positive operator $W$, then $W \backsim W_0$.
\end{Definition}
More precisely, one should call them exposed rays in the convex cone $\mathfrak{L}_1$. If one considers a compact convex set of normalized operators, then exposed points define a subset of extremal ones. One has the celebrated \cite{conv}

\begin{Theorem}[Straszewicz]
Exposed points of a compact convex set  are dense in the set of extremal points.
\end{Theorem}
Interestingly, one proves

\begin{Proposition}[\cite{Marciniak}] All extremal decomposable witnesses are exposed.
\end{Proposition}
For indecomposable witnesses the situation is much more subtle. One has the following necessary condition for a witness to generate an exposed ray

\begin{Proposition} If $\,W$ is an exposed entanglement witness then it has a spanning property.
\end{Proposition}
Hence, a spanning property is sufficient for optimality and necessary for exposedness. Now, we formulate a sufficient condition for exposedness for a class of so called irreducible entanglement witnesses.

\begin{Definition}
A linear map $\Phi : \mathfrak{L}(\HA) \rightarrow \mathfrak{L}(\HB)$ is irreducible if $[\Phi(A),X] =0$ for all $A \in \mathfrak{L}(\HA)$ implies that $X = \lambda \mathbb{I}_B$.
\end{Definition}
Based on the canonical isomorphism introduced in \cite{Pillis} (see Remark \ref{R-Pillis}) one has
\begin{Lemma}
If $\Phi : \mathfrak{L}(\HA) \rightarrow \mathfrak{L}(\HB)$ is a linear map, then
\begin{equation}\label{}
   \sum_{i,j=1}^{d_A} E_{ij} \otimes \Phi(E_{ji}) =  \sum_{\alpha,\beta=1}^{d_B} \Phi^\#(F_{\alpha\beta}) \otimes F_{\beta\alpha}\ ,
\end{equation}
where $F_{\alpha\beta}=|f_\alpha\>\<f_\beta|$ and $\{f_1,\ldots,f_{d_B}\}$ is an arbitrary orthonormal basis in $\HB$.
\end{Lemma}
We call an entanglement witness $W$ irreducible if either $\Phi$ or its dual $\Phi^\#$ is irreducible positive (but not CP) map.
Let us define
\begin{equation*}\label{SPAN-N}
\fl
  N_W = {\rm span}_\mathbb{C}\{\, |\psi\> \otimes |\psi^*\>  \otimes |\phi\> \in \mathcal{H}_A \otimes \HA \otimes \mathcal{H}_B\, |\, \< \psi \otimes \phi |W|\psi \otimes \phi\>  = 0 \, \} \ ,
\end{equation*}
if $\Phi$ is irreducible, and
\begin{equation*}\label{SPAN-N}
\fl
  N^\#_W = {\rm span}_\mathbb{C}\{\, |\psi\> \otimes |\phi\>  \otimes |\phi^*\> \in \mathcal{H}_A \otimes \HB \otimes \mathcal{H}_B\, |\, \< \psi \otimes \phi |W|\psi \otimes \phi\>  = 0 \, \} \ ,
\end{equation*}
if $\Phi^\#$ is irreducible. We say that $W$ possesses a {\em strong spanning property}  if \cite{EXP1,EXP2}
\begin{equation}\label{STRONG}
{\rm dim}\, N_W = (d_A^2-1) d_B \ , \ \ \mbox{or} \ \  {\rm dim}\, N^\#_W = (d_B^2-1) d_A \ .
\end{equation}

\begin{Proposition}[\cite{EXP1}] Any irreducible entanglement witness possessing a strong spanning property is exposed.
\end{Proposition}
Interestingly, one has the following analog of Proposition \ref{Tura}

\begin{Proposition}[\cite{EXP1}]\label{Tura-ex}
For a decomposable irreducible entanglement witness $W$ acting
in $\HA \otimes \HB$ such that $d=\min\{d_A,d_B\} =2$ the following three conditions
are equivalent:
\begin{enumerate}
 \item $W$ has a strong spanning property,
 \item $W$ is exposed,
 \item $W = |\psi\>\<\psi|^\Gamma$.
\end{enumerate}
\end{Proposition}
For more examples of exposed indecomposable entanglement witnesses see sections \ref{S-DIAG} and \ref{S-OPT} (see also  \cite{Marciniak,Majewski,Majewski-Tylec} and the recent review by Kye \cite{Kye-REV}). Examples of extremal entanglement witnesses are analyzed for example in \cite{Sengupta1,Sengupta2}.

\section{Diagonal-type entanglement witnesses in $\mathbb{C}^d \otimes \mathbb{C}^d$}
\label{S-DIAG}

\subsection{General structure}

Let us consider a class of Hermitian operators in $\mathbb{C}^d \otimes \mathbb{C}^d$ defined as follows
\begin{equation}\label{WA!}
  W[A] = \sum_{i,j=1}^d E_{ij} \otimes W_{ij}\ ,
\end{equation}
where $W_{ij} = - E_{ij}$ for $i \neq j$, and the diagonal blocks read
\begin{equation}\label{}
  W_{ii} = \sum_{k=1}^d a_{ik} E_{kk} \ ,
\end{equation}
with $a_{ik}\geq 0$ being the matrix elements of $d \times d$ matrix $A$. Hence off diagonal blocks $W_{ij}$ are fixed and all diagonal blocks $W_{ii}$ are represented by diagonal $d \times d$ matrices with $a_{ij}$ entries, i.e. $(W_{ii})_{jj} = a_{ij}$.   For obvious reason we shall call $W[A]$ a {\em diagonal type operator}. As we shall see several important examples of witnesses belong to class (\ref{WA!}).
Interestingly, a  diagonal-type operator possesses the following local symmetry
\begin{equation}\label{}
  U \otimes U W[A] U^\dagger \otimes U^\dagger = W[A]\ ,
\end{equation}
where $U$ is a unitary operator defined by
\begin{equation}\label{}
  U = \sum_{k=1}^d e^{i\lambda_k} E_{kk}\ ,
\end{equation}
with real $\lambda_k$, that is, $W[A]$ is invariant under maximal commutative subgroup (a $d$-dimensional torus) of $U(d)$.

\begin{Theorem}[\cite{DC1}] A diagonal-type operator $W[A]$ is block positive if and only if for all vectors $x \in \mathbb{C}^d$
\begin{equation}\label{W-A}
  \sum_{i=1}^d \frac{|x_i|^2}{B_i(x)} \leq 1\ ,
\end{equation}
where $B_i(x) = |x_i|^2 + \sum_{j=1}^d a_{ij} |x_j|^2$. Moreover, $W[A] \geq 0$ if and only if the matrix $D$ such that $D_{ij} = -1$ for $i \neq j$ and $D_{ii} = a_{ii}$ is positive semi-definite.
\end{Theorem}

\begin{Example}
For  $d=2$ one finds
\begin{equation}\label{}
  W[A]\, =\, %\frac{N_{abc}}{3}\,
  \left( \begin{array}{cc|cc} a_{11} & . & . & -1 \\ . & a_{12} & . & . \\ \hline . & . & a_{21} & . \\ -1 & . & . & a_{22} \end{array} \right)\ ,
\end{equation}
where to make the picture more transparent we replaced zeros by dots.
The matrices $A$ and $D$ are defined as follows
\begin{equation}\label{}
A \, =\, %\frac{N_{abc}}{3}\,
  \left( \begin{array}{cc} a_{11} & a_{12}  \\ a_{21} & a_{22}  \end{array} \right)\ , \ \ \   D \, =\, %\frac{N_{abc}}{3}\,
  \left( \begin{array}{cc} a_{11} & -1  \\ -1 & a_{22}  \end{array} \right)\ ,
\end{equation}
and hence the condition (\ref{W-A}) reads
\begin{equation}\label{}
  \frac{|x_1|^2}{(a_{11} + 1)|x_1|^2 + a_{12}|x_2|^2} +  \frac{|x_2|^2}{(a_{22} + 1)|x_2|^2 + a_{21}|x_1|^2} \leq 1 \ .
\end{equation}
Taking $|x_1|=|x_2|$ one finds
\begin{equation}\label{aa}
 a_{11}a_{22} + a_{12}a_{21} \geq 1 - a_{11}a_{21} - a_{22}a_{12}\ .
\end{equation}
Let $a_{11}a_{22} = p^2$ and $a_{12}a_{21}=q^2$.  One has
$$ a_{11}a_{22} + a_{12}a_{21} = p^2 + q^2 = (p+q)^2 - 2 pq \ . $$
Now, $2pq = 2 \sqrt{a_{11}a_{21}a_{22}a_{12} } \leq a_{11}a_{21} + a_{22}a_{12}$ and hence
$$ a_{11}a_{22} + a_{12}a_{21} \geq  (p+q)^2 - a_{11}a_{21} - a_{22}a_{12} \ , $$
which shows that (\ref{aa}) is equivalent to $p+q\geq 1$. This condition is equivalent to block-positivity of $W[A]$ since in this case all block-positive operators are decomposable.  Indeed, one has
\begin{equation*}\label{}
\fl
  \left( \begin{array}{cc|cc} a_{11} & . & . & -1 \\ . & a_{12} & . & . \\ \hline . & . & a_{21} & . \\ -1 & . & . & a_{22} \end{array} \right) =
  \left( \begin{array}{cc|cc} a_{11} & . & . & -1+q \\ . & a_{12} & . & . \\ \hline . & . & a_{21} & . \\ -1+q & . & . & a_{22} \end{array} \right) + \left( \begin{array}{cc|cc} a_{11} & . & . & . \\ . & a_{12} & -q & . \\ \hline . & -q & a_{21} & . \\ . & . & . & a_{22} \end{array} \right)^\Gamma\ ,
\end{equation*}
which shows that $W[A]$ is a decomposable entanglement witness whenever $a_{11}a_{22} = p^2 \leq 1$, i.e. $D$ is not a positive matrix.
\end{Example}

%\begin{Remark}
For $d \geq 3$ condition (\ref{W-A}) is highly nontrivial. Only some special cases were worked out. Consider the matrix $A$ defined by
\begin{equation}\label{}
  a_{ii} = a\ , \ \ \ a_{ij} = \delta_{i-1,j}\, c_j \ , \ \ i\neq j\ .
\end{equation}
The structure of $A$ for $d=3,4,5$ reads as follows
\begin{equation*}\label{}
  \left( \begin{array}{ccc} a & . & c_3  \\ c_1 & a & . \\ . & c_2 & a  \end{array} \right)\  , \ \ \
  \left( \begin{array}{cccc} a & . & . & c_4  \\ c_1 & a & . & . \\ . & c_2 & a & . \\ . & . & c_3 & a   \end{array} \right)\ , \ \ \
  \left( \begin{array}{ccccc} a & . & . & . & c_5  \\ c_1 & a & . & . & . \\ . & c_2 & a & . & . \\ . & . & c_3 & a & . \\ . & . & . & c_4 & a\end{array} \right)\ .
\end{equation*}
One has therefore
\begin{equation}\label{or}
  W_{ii} = a E_{ii} + c_{i-1} E_{i-1,i-i} \ ,
\end{equation}
for $i=1,\ldots,d$.
\begin{Theorem}[\cite{Ha2}] Let $ a,c_1\ldots,c_d>0$. $W[a;c_1,\ldots,c_d]$ is an entanglement witness if and only if
\begin{enumerate}
\item $  d-1 > a \geq d-2$,
\item $  (c_1 \cdot \ldots \cdot c_d)^{1/d}  \geq d-1- a$.
\end{enumerate}
\end{Theorem}
Interestingly, $W[a;c_1,\ldots,c_d]$ provides an atomic entanglement witness \cite{Ha2}.
\begin{Proposition}[\cite{Osaka1,Osaka2}] $W[1;c_1,c_2,c_3]$ is an extremal entanglement witness if $\, c_1\,  c_2\,  c_3 = 1$.
\end{Proposition}
The above class of witnesses was recently analyzed in \cite{Li-Wu}: for any  permutation $\sigma$ from the symmetric group $S_d$ one defines \cite{Li-Wu} a diagonal-type operator $W_\sigma[a;c_1,\ldots,c_d]$ by
\begin{equation}\label{}
  W_{ii} = a E_{ii} + c_{i-1} E_{\sigma(i),\sigma(i)} \ .
\end{equation}
The original formula (\ref{or}) is recovered for $\sigma(i) = i-1$.

\begin{Proposition}[\cite{Li-Wu}] Let $a,c_1,\ldots,c_d > 0$ and $\sigma \in S_d$:

\begin{enumerate}
\item if $W[a;c_1,\ldots,c_d]$ is block positive, then $W_\sigma[a;c_1,\ldots,c_d]$ is block positive,
\item if $\sigma$ is a cycle of length $d$, then  $W_\sigma[a;c_1,\ldots,c_d]$ is block positive iff $W[a;c_1,\ldots,c_d]$ is block positive.
\end{enumerate}
\end{Proposition}
Now, let $\sigma = \sigma_1 \circ \ldots \circ \sigma_r$ be a unique decomposition of $\sigma$ into disjoint cycles. Denoting by $l(\sigma_k)$ a length of a cycle $\sigma_k$ one introduces
$$  l_{\rm max}(\sigma) = \max \{ l(\sigma_1),\ldots,l(\sigma_r)\} \ , \ \ \
l_{\rm min}(\sigma) = \min \{ l(\sigma_1),\ldots,l(\sigma_r)\} \ . $$
Let $\tau^d_k \in S_d$ denote a cycle defined by
$   \tau^d_k(i) = i+k  , \ ({\rm mod}\ d)$.

\begin{Proposition} Let $\sigma \in S_d$ and $d \geq 3$.

\begin{itemize}
\item If $l_{\rm min}(\sigma) \geq 3$ and  $0 < c \leq \frac{d}{l_{\rm max}(\sigma)}$, then $W_\sigma[d-c,c,\ldots,c]$ is atomic.

\item For each $k \in \{1,\ldots,d-1\}$,  if $k \neq \frac d2 $ when $d$ is even, then $W_{\tau^d_k}[d-2,1,\ldots,1]$ is atomic.

\end{itemize}

\end{Proposition}
A wide class of matrices $A$ giving rise to diagonal-type entanglement witnesses was proposed in \cite{Kossak-kule}:
 let us define a set of Hermitian traceless matrices
\begin{equation}\label{}
    F_\ell  = \frac{1}{\sqrt{\ell(\ell +1)}} \Big( \sum_{k=1}^{\ell-1} E_{kk} - \ell E_{\ell\ell} \Big) \ ,\ \ \ \ \ell = 1,\ldots,d-1\ .
\end{equation}
One defines a real $d \times n$ matrix
\begin{equation}\label{a-R}
    a_{ij} = \frac{d-1}{d} + \sum_{\alpha,\beta=1}^{d-1}  \< e_i|F_\alpha|e_i\> R_{\alpha\beta} \<e_j|F_\beta|e_j\> \ ,
\end{equation}
where $R_{\alpha\beta}$ is an orthogonal $(d-1)\times (d-1)$ orthogonal matrix. Due to the fact that $F_\alpha$ is traceless
for $\alpha=1,\ldots,d-1$, one finds
\begin{equation}\label{doubly}
    \sum_{i=1}^{d-1} a_{ij} = \sum_{j=1}^{d-1} a_{ij} = d-1\ ,
\end{equation}
Moreover, it turns out \cite{Kossak-kule} that $a_{ij} \geq 0$ and hence $\widetilde{A} = \frac{1}{d-1} A$ defines a doubly stochastic matrix. One proves

\begin{Proposition}[\cite{Kossak-kule}] \label{KULE}
For any orthogonal matrix $R_{\alpha\beta}$ a diagonal-type operator $W[A]$, where $A$ is defined by (\ref{a-R}) is block-positive.
\end{Proposition}
For more examples of diagonal-type witnesses see also \cite{DC1,DC-kule,Filip2}.

%\end{Example}

\subsection{Generalized Choi witnesses}

Now, we analyze important class of diagonal-type entanglement witnesses which provide the generalization of the EW proposed by Choi \cite{Choi}.
The analysis of $W[A]$ simplifies if $A$ is a circulant matrix, i.e. $a_{ij} = \alpha_{i-j}$, where one adds modulo $d$ and $\alpha_k \geq 0$ for $k=0,1,\ldots,d-1$. In this case (\ref{W-A}) reduces to so called circular inequalities \cite{Yamagami}
\begin{equation}\label{W-A-c}
  \sum_{i=1}^d \frac{|x_i|^2}{(\alpha_0 + 1)|x_i|^2 + \sum_{k=1}^{d-1} \alpha_k |x_{i+k}|^2 } \leq 1\ .
\end{equation}
In particular taking $|x_1| = \ldots = |x_d|$ one finds
\begin{equation}\label{d-1}
  \alpha_0 + \alpha_1 + \ldots + \alpha_{d-1} \geq d-1\ .
\end{equation}
This condition is necessary but not sufficient for $W[A]$ to be an entanglement witness. Actually, it is sufficient only for $d=2$. If $A$ is a circulant matrix we  denote $W[A] = W[\alpha_0,\ldots,\alpha_{d-1}]$.

\begin{Cor} $W[a,b]$ defined by
\begin{equation}\label{}
  W[a,b]\, =\, %\frac{N_{abc}}{3}\,
  \left( \begin{array}{cc|cc} a & . & . & -1 \\ . & b & . & . \\ \hline . & . & b & . \\ -1 & . & . & a \end{array} \right)\ ,
\end{equation}
is an entanglement witness if and only if $a < 1$ and $a+b \geq 1$.
\end{Cor}
Note that $W[0,1]$ corresponds to the reduction map in $M_2(\mathbb{C})$ defined by $R_2(X) = \mathbb{I}_2 {\rm tr}X - X$.
For $d=3$  circular inequalities reduce to
\begin{equation}\label{circ-3}
\fl
  \frac{t_1}{(a+1) t_1 + b\, t_2 + c\, t_3} +  \frac{t_2}{(a+1)t_2 + b\, t_3 + c\, t_1} + \frac{ t_3}{(a+1) t_3 + b\, t_1 + c\, t_2} \leq 1\ ,
\end{equation}
where we introduced $t_k = |x_k|^2$, $a=\alpha_0$, $b=\alpha_1$ and $c=\alpha_2$. Hence, inequalities (\ref{circ-3}) are satisfied for all  $t_k \geq 0$ if and only if the following operator
\begin{equation}\label{W-abc}
% \hspace*{-.1cm}
  W[a,b,c]\, =\, %\frac{N_{abc}}{3}\,
  \left( \begin{array}{ccc|ccc|ccc}
    a & \cdot & \cdot & \cdot & -1 & \cdot & \cdot & \cdot & -1 \\
    \cdot& b & \cdot & \cdot & \cdot& \cdot & \cdot & \cdot & \cdot  \\
    \cdot& \cdot & c & \cdot & \cdot & \cdot & \cdot & \cdot &\cdot   \\ \hline
    \cdot & \cdot & \cdot & c & \cdot & \cdot & \cdot & \cdot & \cdot \\
    -1 & \cdot & \cdot & \cdot & a & \cdot & \cdot & \cdot & -1  \\
    \cdot& \cdot & \cdot & \cdot & \cdot & b & \cdot & \cdot & \cdot  \\ \hline
    \cdot & \cdot & \cdot & \cdot& \cdot & \cdot & b & \cdot & \cdot \\
    \cdot& \cdot & \cdot & \cdot & \cdot& \cdot & \cdot & c & \cdot  \\
    -1 & \cdot& \cdot & \cdot & -1 & \cdot& \cdot & \cdot & a
     \end{array} \right)\ ,
\end{equation}
is block positive. One proves the following

\begin{Theorem}[\cite{Cho-abc}]  \label{TH-korea}
An operator $W[a,b,c]$ is an entanglement witness if and only if
\begin{enumerate}
\item $0 \leq a < 2\ $,
\item $ a+b+c \geq 2\ $,
\item if $a \leq 1\ $, then $ \ bc \geq (1-a)^2$.
\end{enumerate}
Moreover, being being an entanglement witness  it is indecomposable if and only if
\begin{equation}\label{ind}
  4bc < (2-a)^2\ .
\end{equation}
$W[a,b,c]$ is a 3-Schmidt witness if and only if $\, 2> a \geq 1$ and $bc \geq (2-a)(b+c)$.
\end{Theorem}
Note, that if $W[a,b,c]$ is a 3-Schmidt witness, i.e. $W[a,b,c] \in \mathfrak{L}_2 - \mathfrak{L}_3$,  then it is necessarily decomposable. One proves \cite{Ha-RIMS} that any indecomposable $W[a,b,c]$ is necessarily atomic.
This class of witnesses corresponds to a family of positive maps  $\Phi[a,b,c] : M_3(\mathbb{C}) \rightarrow M_3(\mathbb{C})$
\begin{eqnarray}\label{Choi-map}
    \Phi[a,b,c](E_{ii}) &=& \frac 1N \sum_{i,j=1}^3 A_{ij}[a,b,c] E_{jj} \ , \\
\Phi[a,b,c](E_{ij}) &=& - \frac{1}{N} \ E_{ij} \ , \ \ i\neq
j\ ,
\end{eqnarray}
where the $3 \times 3$ matrix $A[a,b,c]$ is defined by
\begin{equation}\label{}
    A[a,b,c] =  \left( \begin{array}{ccc} a & b &
    c \\ c & a & b \\
    b & c & a  \end{array} \right) \ .
\end{equation}
The normalization factor $N = a+b+c$ guaranties that the map is unital, i.e. $\Phi[a,b,c](\mathbb{I}_3) = \mathbb{I}_3$. This class of maps contains well known examples: the Choi  indecomposable map $\Phi[1,1,0]$ and decomposable reduction map $\Phi[0,1,1]$ which may be rewritten as follows
\begin{equation}\label{R}
    \Phi[0,1,1](X) = \frac 12 ( \mathbb{I}_3\, {\rm tr} X - X)\ .
\end{equation}
Note, that the dual map to $\Phi[a,b,c]$ reads $\Phi^\#[a,b,c]= \Phi[a,c,b]$. Hence, the reduction map is self-dual and $\Phi^\#[1,1,0] = \Phi[1,0,1]$.
Choi maps $\Phi[1,1,0]$ and $\Phi[1,0,1]$ provided first examples of positive indecomposable maps \cite{Choi}.

%Note, that for $a\geq 2$ one has $W[a,b,c] \geq 0$.   It is clear that if $W[a,b,c]$ is an EW and $a'\geq a$, $b'\geq b$, $c'\geq c$, then $W[a',b',c']$ is %block-positive and provides  EW whenever $a'<2$. Note that if $\{a_1,b_1,c_1\}$ and $\{a_2,b_2,c_2\}$ satisfy {\it 1}, {\it  2} and {\it 3}, then
%\begin{equation}\label{}
%  W[a,b,c] := p_1 W[a_1,b_1,c_1] +  p_2 W[a_2,b_2,c_2]\ ,
%\end{equation}
%where $p_1+p_2=1$ and $p_1,p_2\geq 0$ defines EW  with
%$$ a= p_1 a_1 + p_2 a_2\ , \ \  b= p_1 b_1 + p_2 b_2\ , \ \ c= p_1 c_1 + p_2 c_2\ . $$
%Hence, entanglement witnesses $W[a,b,c]$ define a convex cone.

It should be clear that to analyze a set of admissible $\{a,b,c\}$ giving rise to optimal entanglement witnesses it is enough to consider a 2-dimensional subset satisfying
\begin{equation}\label{abc=2}
    a+b+c=2 \ .
\end{equation}
%Witnesses satisfying (\ref{abc=2})  belong to the boundary of the general class satisfying $a+b+c \geq 2$.
Assuming (\ref{abc=2}) a family of entanglement witnesses $W[a,b,c]$ is essentially parameterized by two parameters $b$ and $c$. The boundary of this set on the $bc$-plane is governed by condition {\em 3.} of Theorem \ref{TH-korea}: it gives rise to a part of an  ellipse $bc = (1+b-c)^2$ (see \cite{Filip1} for details).  The properties of $W[2-b-c,b,c]$ belonging to the boundary set may be summarized as follows

%\begin{figure}[htp] \label{fig}
% \centering
%\includegraphics[scale=0.65]{elipsa_2}
%\includegraphics[scale=0.65]{bc_3}
%\caption{A convex set of entanglement witnesses $W[2-b-c,b,c]$. A line $b=c$ corresponds to decomposable witnesses. Special points: (i) and (ii) Choi %witnesses, (iii) reduction witness, (v) defined a positive operator $W[2,0,0]$, (iv) decomposable witness with $b=c=1/3$.}
%\end{figure}

\begin{enumerate}

%\item $W$ is atomic for $b \neq c$,

\item apart from the Choi witnesses $W[1,1,0]$ and $W[1,0,1]$ the remaining witnesses have co-spanning property \cite{Ha-opt,nasza-opt} and hence they are nd-optimal (or bi-optimal according to \cite{cospan}),

\item all witnesses apart from $W[0,1,1]$ are extremal (the extremality of $W[1,1,0]$ and $W[1,0,1]$ was already proved by Choi and Lam \cite{Choi-Lam}. Clearly, being extremal and indecoposable they are nd-optimal),

\item Choi witnesses $W[1,1,0]$ and $W[1,0,1]$ are extremal but not exposed,

\item all remaining extremal witnesses are exposed \cite{Kye-EXP1,Kye-EXP2}. None of them possesses a strong spanning property.

\item all optimal witnesses satisfy SPA conjecture \cite{Filip1}.

\end{enumerate}

\begin{Remark} Interestingly, a boundary elements from the class $W[2-b-c,b,c]$ may be recovered from Proposition \ref{KULE}.
%Indeed, it is not difficult to construct an orthonormal basis $F_\alpha$ in $M_d(\mathbb{C})$. One may take for example the generalized %Gell-Mann matrices defined as follows: let $|1\>,\ldots,|d\>$ be an orthonormal basis in $\mathbb{C}^n$ and define
%\begin{eqnarray*}
%d_l &=& \frac{1}{\sqrt{l(l+1)}}\Big( \sum_{k=1}^l E_{kk} -l E_{ll}\Big)\ ,\ \ \ l=1,\ldots,d-1 \\
%u_{kl} &=&\frac{1}{\sqrt{2}}(E_{kl}+E_{lk})\ , \\
%v_{kl} &=&\frac{-i}{\sqrt{2}}(E_{kl}-E_{lk})\ ,
%\end{eqnarray*}
%for $k<l$. It is easy to see that $d^2$ Hermitian matrices $(F_0,d_l,u_{kl},v_{kl})$ define a proper orthonormal basis in $M_d(\mathbb{C})$. Now, let $d=3$ and consider the following orthogonal matrix $R \in O(8)$:
%\begin{equation}\label{}
%   R = \left(\begin{array}{c|c} T & 0 \\ \hline 0 & - \mathbb{I}_6 \end{array} \right)\ ,
%\end{equation}
%where $T \in O(2)$ is a proper rotation defined by
Taking the following orthogonal matrix
\begin{equation}\label{T}
    R(\alpha) = \left(\begin{array}{cc} \cos\alpha & -\sin\alpha  \\ \sin\alpha & \cos\alpha \end{array} \right) \ ,
\end{equation}
for $\alpha \in [0,2\pi)$, one finds that the elements $a_{ij}$ defined by (\ref{a-R})give rise to a circulant matrix, that is, $W[A] = W[a(\alpha),b(\alpha),c(\alpha)]$, where the $\alpha$-dependent coefficients $a,b,c$ are defined as follows \cite{Kossak-kule}:
\begin{eqnarray}   \label{abc}
\fl
 b(\alpha) =\frac{2}{3}\left(1-\frac{1}{2}\cos\alpha-\frac{\sqrt{3}}{2}\sin\alpha\right) \ , \  \
c(\alpha) =\frac{2}{3}\left(1-\frac{1}{2}\cos\alpha+\frac{\sqrt{3}}{2}\sin\alpha\right)\ , \nonumber
\end{eqnarray}
and  $ a(\alpha) = 2 - b(\alpha)- c(\alpha)= \frac{2}{3}\,(1+\cos\alpha)$. It is easy to show that
\begin{equation}\label{ABC}
    b(\alpha)c(\alpha) = [1-a(\alpha)]^2\ ,
\end{equation}
for $ \alpha \in [0,2\pi)$. Hence, this class satisfies conditions of Theorem \ref{TH-korea}. The parameter $ \alpha \in [0,2\pi)$ provides a natural parametrization of a boundary  ellipse \cite{Filip1}.

\end{Remark}

\begin{Remark}
For $d \geq 4$ the general conditions for circulant Choi-like witness are not known. Such witnesses were recently analyzed in \cite{Filip2}.
\end{Remark}
Consider now a  discrete family of Choi-like witnesses \cite{Ha-RIMS,Tanahashi,Osaka3} constructed as follows:
\begin{equation}\label{Wdk}
  W_{11} = (d-k) E_{11} + E_{22} + \ldots + E_{k-1,k-1} \ ,
\end{equation}
and
\begin{equation}\label{}
  W_{jj} = \mathbf{S}^j W_{11} \mathbf{S}^{\dagger j}\ ,
\end{equation}
where $\mathbf{S} : \mathbb{C}^d \rightarrow \mathbb{C}^d$ is a unitary shift defined by
\begin{equation}\label{}
  \mathbf{S} e_k = e_{k+1} \ , \ \ \ i=k, \ldots , d\ .
\end{equation}
It means that a circulant matrix $A$ is defined by
\begin{equation}\label{}
  \alpha_0 = d-k, \ \alpha_1 = \ldots =\alpha_{k-1} = 1\ ,
\end{equation}
and the rest $\alpha_k= \ldots = \alpha_{d-1}=0$. One has $\alpha_0 + \ldots + \alpha_{d-1}= d-1$ and hence it is compatible with (\ref{d-1}).

\begin{Proposition}[\cite{Ha-RIMS,Tanahashi,Osaka3}] $W_{d,k}$ is block-positive  for all $k=1,\ldots,d$. Moreover

\begin{enumerate}

\item $W_{d,1} \geq 0$ and $W_{d,d}^\Gamma \geq 0$,

\item $W_{d,k}$ is an indecomposable entanglement witness for $k=2,\ldots,d-1$.

\end{enumerate}

\end{Proposition}

%\end{Example}
For more examples of diagonal-type Choi-like witnesses see also \cite{JAFA}.

%\section{Robertson-like entanglement witnesses in $\mathbb{C}^{2N} \otimes \mathbb{C}^{2N}$}

\section{A class of optimal entanglement witnesses in $\mathbb{C}^{2N} \otimes \mathbb{C}^{2N}$}   \label{S-OPT}

In this section we provide  several examples of optimal entanglement witnesses in  $\mathbb{C}^{2N} \otimes \mathbb{C}^{2N}$ which are not of the diagonal type.

\subsection{Robertson map  in $M_4(\mathbb{C})$}

We start with the construction of a positive map in $M_4(\mathbb{C})$ proposed by Robertson \cite{Robertson1}. Note that any operator $X \in M_4(\mathbb{C})$ may be written in a block form
\begin{equation}\label{}
  X = \left( \begin{array}{c|c} X_{11} & X_{12} \\ \hline X_{21} & X_{22} \end{array} \right)\ ,
\end{equation}
where $X_{ij} \in M_2(\mathbb{C})$ for $i,j=1,2$. Now, following Robertson \cite{Robertson1} one defines a map $\Phi_4 : M_4(\mathbb{C}) \rightarrow M_4(\mathbb{C})$ by
\begin{equation}\label{R4}
    \Phi_4\left( \begin{array}{c|c} X_{11} & X_{12} \\ \hline X_{21} & X_{22} \end{array}
\right) = \frac 12 \left( \begin{array}{c|c} \mathbb{I}_2\,
\mbox{Tr} X_{22} &  -[X_{12} + R_2(X_{12}^{\rm t})] \\ \hline  -[X_{21} +
R_2(X_{21}^{\rm t})] & \mathbb{I}_2\, \mbox{Tr} X_{11}
\end{array} \right) \ ,
\end{equation}
where $R_2$ denotes a reduction map in $M_2(\mathbb{C})$:
\begin{equation}\label{}
  R_2\left( \begin{array}{cc} x_{11} & x_{12} \\ x_{21} & x_{22} \end{array}
\right) =  \left( \begin{array}{cc} x_{22} & -x_{12} \\ -x_{21} & x_{11} \end{array}
\right)\ .
\end{equation}
 The normalization factor guaranties that $\Phi_4$ is unital and trace-preserving.
One easily finds the action of $\Phi_4$ on the matrix units $E_{ij}$ in $M_4(\mathbb{C})$:
$$  \Phi_4(E_{11}) = \Phi_4(E_{22}) = \frac 12 (E_{33} + E_{44})\ , \ \ \  \Phi_4(E_{22}) = \Phi_4(E_{33}) = \frac 12 (E_{11} + E_{22})\ , $$
$$  \Phi_4(E_{13}) = - \frac 12 (E_{13} + E_{24}) \ , \ \ \  \Phi_4(E_{24}) = - \frac 12 (E_{24} + E_{31}) \ , $$
$$  \Phi_4(E_{14}) = - \frac 12 (E_{14} - E_{32}) \ , \ \ \  \Phi_4(E_{23}) = - \frac 12 (E_{23} + E_{41}) \ , $$
and the remaining $\Phi_4(E_{12}) =  \Phi_4(E_{34}) = 0$.

\begin{Proposition} The Robertson map $\Phi_4$ enjoys the following properties:

\begin{enumerate}

\item it is positive extremal (and hence optimal) \cite{Robertson1},

\item it is atomic \cite{Ha-RIMS} and hence indecomposable (see also \cite{DC3}),

\item it is exposed \cite{DC14}.

\item satisfies SPA conjecture \cite{DC7}.

\end{enumerate}

\end{Proposition}
The corresponding entanglement witness $W_4 = 2\sum_{i,j=1}^4 E_{ij} \otimes \Phi_4(E_{ij})$ displays the following matrix structure

\begin{equation}\label{}
\fl
 W_4 = \left( \begin{array}{cccc|cccc|cccc|cccc}
 \cdot& \cdot& \cdot& \cdot& \cdot& \cdot& \cdot& \cdot& \cdot& \cdot& -1& \cdot& \cdot& \cdot& \cdot& -1\\
 \cdot& \cdot& \cdot& \cdot& \cdot& \cdot& \cdot& \cdot& \cdot& \cdot& \cdot& \cdot& \cdot& \cdot& \cdot& \cdot\\
 \cdot& \cdot& 1& \cdot& \cdot& \cdot& \cdot& \cdot& \cdot& \cdot& \cdot& \cdot& \cdot& 1& \cdot& \cdot\\
 \cdot& \cdot& \cdot& 1& \cdot& \cdot& \cdot& \cdot& \cdot& 1& \cdot& \cdot& \cdot& \cdot& \cdot& \cdot  \\ \hline
 \cdot& \cdot& \cdot& \cdot& \cdot& \cdot& \cdot& \cdot& \cdot& \cdot& \cdot& \cdot& \cdot& \cdot& \cdot& \cdot \\
 \cdot& \cdot& \cdot& \cdot& \cdot& \cdot& \cdot& \cdot& \cdot& \cdot& -1& \cdot& \cdot& \cdot& \cdot& -1 \\
 \cdot& \cdot& \cdot& \cdot& \cdot& \cdot& 1& \cdot& \cdot& \cdot& \cdot& \cdot& -1& \cdot& \cdot& \cdot \\
 \cdot& \cdot& \cdot& \cdot& \cdot& \cdot& \cdot& 1& 1& \cdot& \cdot& \cdot& \cdot& \cdot& \cdot& \cdot  \\ \hline
 \cdot& \cdot& \cdot& \cdot& \cdot& \cdot& \cdot& 1& 1& \cdot& \cdot& \cdot& \cdot& \cdot& \cdot& \cdot \\
 \cdot& \cdot& \cdot& -1& \cdot& \cdot& \cdot& \cdot& \cdot& 1& \cdot& \cdot& \cdot& \cdot& \cdot& \cdot \\
 -1& \cdot& \cdot& \cdot& \cdot& -1& \cdot& \cdot& \cdot& \cdot& \cdot& \cdot& \cdot& \cdot& \cdot& \cdot \\
 \cdot& \cdot& \cdot& \cdot& \cdot& \cdot& \cdot& \cdot& \cdot& \cdot& \cdot& \cdot& \cdot& \cdot& \cdot& \cdot \\ \hline
 \cdot& \cdot& \cdot& \cdot& \cdot& \cdot& -1& \cdot& \cdot& \cdot& \cdot& \cdot& 1& \cdot& \cdot& \cdot \\
 \cdot& \cdot& 1& \cdot& \cdot& \cdot& \cdot& \cdot& \cdot& \cdot& \cdot& \cdot& \cdot& 1& \cdot& \cdot \\
 \cdot& \cdot& \cdot& \cdot& \cdot& \cdot& \cdot& \cdot& \cdot& \cdot& \cdot& \cdot& \cdot& \cdot& \cdot& \cdot \\
 -1& \cdot& \cdot& \cdot& \cdot& -1& \cdot& \cdot& \cdot& \cdot& \cdot& \cdot& \cdot& \cdot& \cdot& \cdot \end{array}
 \right)\ ,
\end{equation}
and clearly it is not of the diagonal type.

\subsection{Breuer-Hall maps  in $M_{2N}(\mathbb{C})$}

Consider a class of linear maps $\Phi_U :  M_{2N}(\mathbb{C}) \rightarrow
M_{2N}(\mathbb{C})$ defined as follows \cite{Breuer,Hall}
\begin{equation}\label{B-H}
\Phi_U(X) =  \frac{1}{2(N-1)} \Big[  \mathbb{I}_{2N}\, {\rm tr} X - X - UX^{\rm t} U^\dagger \Big]  \ ,
\end{equation}
where $U$ is an antisymmetric  unitary  matrix in $\mathbb{C}^{2N}$. The normalization factor guaranties that $\Phi_U$ is unital and trace-preserving.
 The characteristic feature of these maps is that for any rank one
projector $P$ its image under $\Phi_U$ reads
\begin{equation}\label{}
    \Phi_U(P) = \frac{1}{2(N-1)} [\mathbb{I}_{2N} - P - Q] \ ,
\end{equation}
where $Q$ is again rank one projector satisfying $PQ=0$, that is, if $P=|\psi\>\<\psi|$, then $Q=  |\widetilde{\psi}\>\<\widetilde{\psi}|$, where $\widetilde{\psi} = U|\psi\>$. Antisymmetry of $U$ guaranties that $\<{\psi}|\widetilde{\psi}\> = \<\psi|U|\psi\> = 0$.     Hence
$\mathbb{I}_{2N} - P - Q$ is a projector onto the subspace orthogonal to $|\psi\>$ and $|\widetilde{\psi}\>$ which proves positivity of $\Phi_U$.

\begin{Proposition} The Breuer-Hall map $\Phi_U$ is

\begin{enumerate}

\item atomic \cite{DC3} and hence indecomposable \cite{Breuer,Hall},

\item nd-optimal \cite{Breuer},

\item exposed \cite{DC14} for $N=2$.

\end{enumerate}

\end{Proposition}

\begin{Remark} If $N=2$ and
\begin{equation}\label{}
  U =  U_0 = i\, \mathbb{I}_2 \otimes \sigma_2 \ = \  \left( \begin{array}{cc|cc}
    0 & 1 & 0 & 0 \\
    -1& 0 & 0 & 0 \\ \hline
    0 & 0 & 0 & 1 \\
    0 & 0 & -1& 0 \end{array} \right)\ ,
\end{equation}
then $\Phi_{U_0}$ reproduces the Robertson map $\Phi_4$. Hence, taking
\begin{equation}\label{}
  U_0 = i\, \mathbb{I}_N \otimes \sigma_2 \ ,
\end{equation}
one obtains a natural generalization of the Robertson map from $M_4(\mathbb{C})$ to $M_{2N}(\mathbb{C})$ \cite{DC7,DC9,DC10}: representing $X \in M_{2N}(\mathbb{C})$ as an $N \times N$ matrix with $2\times 2$ blocks $X_{ij}$ one finds
\begin{eqnarray} \fl
\Phi_{U_0}\left(\begin{array}{c|c|c|c}
X_{11} & X_{12} & \cdots & X_{1k}\\
\hline X_{21} & X_{22} & \cdots & X_{2k}\\
\hline \vdots & \vdots & \ddots & \vdots\\
\hline X_{k1} & X_{k2} & \cdots & X_{kk}\end{array}\right)=\frac{1}{2(k-1)}\left(\begin{array}{c|c|c|c}
A_{1} & -B_{12} & \cdots & -B_{1k}\\
\hline -B_{21} & A_{2} & \cdots & -B_{2k}\\
\hline \vdots & \vdots & \ddots & \vdots\\
\hline -B_{k1} & -B_{k2} & \cdots & A_{k}\end{array}\right)\ ,\label{100}
\end{eqnarray}
where
\begin{equation}
A_{k}=\mathbb{I}_{2}(\mathrm{Tr}X-\mathrm{Tr}X_{kk})\ ,\label{Ak}
\end{equation}
and
\begin{equation}
B_{kl}=X_{kl}+R_{2}(X^{\rm t}_{kl})\ .\label{Bkl}
\end{equation}

\end{Remark}

\subsection{Generalizations of the Robertson map}

In this section we present several generalizations of the Robertson map.

Formula (\ref{R4}) may be generalized as follows \cite{DC7}:  representing $X \in M_{2N}(\mathbb{C})$ as  $2 \times 2$ matrix with $N\times N$ blocks $X_{ij}$ one defines a linear map $\Phi_{2N} : M_{2N}(\mathbb{C}) \rightarrow M_{2N}(\mathbb{C})$

\begin{equation}\label{RN}
\fl
    \Phi_{2N}\left( \begin{array}{c|c} X_{11} & X_{12} \\ \hline X_{21} & X_{22} \end{array}
\right) = \frac 1N \left( \begin{array}{c|c} \mathbb{I}_N\,
\mbox{Tr} X_{22} &  -[X_{12} + R_N(X_{12}^{\rm t})] \\ \hline  -[X_{21} +
R_N(X_{21}^{\rm t})] & \mathbb{I}_N\, \mbox{Tr} X_{11}
\end{array} \right) \ ,
\end{equation}
where $R_N$ denotes a reduction map in $M_N(\mathbb{C})$, i.e. $R_N(X) = \mathbb{I}_N {\rm tr}X - X$. For $N=2$ it reproduces the Robertson map (\ref{R4}).

\begin{Proposition}[\cite{DC7}] A map $\Phi_{2N}$ is positive, indecomposable and optimal.
\end{Proposition}
Recently, it was shown \cite{EXP2}, that $\Phi_{2N}$ is exposed and hence extremal.

The second generalization uses well known representation of $R_2$
\begin{equation}\label{}
  R_2(X) = \sigma_y X^{\rm t} \sigma_y\ ,
\end{equation}
which provides Kraus representation of a completely positive map $R_2 \circ {\rm T}$. The Pauli matrix $\sigma_2$ is unitary and antisymmetric. Now, we replace $R_{2N}$ by $U X U^\dagger$, with $U$ being unitary antisymmetric matrix in $\mathbb{C}^{2N}$, that is,
we define a linear map $\Psi_{4N} : M_{4N}(\mathbb{C}) \rightarrow M_{4N}(\mathbb{C})$

\begin{equation}\label{RN}
\fl
    \Psi^U_{4N}\left( \begin{array}{c|c} X_{11} & X_{12} \\ \hline X_{21} & X_{22} \end{array}
\right) = \frac{1}{2N} \left( \begin{array}{c|c} \mathbb{I}_{2N}\,
\mbox{Tr} X_{22} &  -[X_{12} + UX_{12}U^\dagger] \\ \hline  -[X_{21} +
U X_{21} U^\dagger] & \mathbb{I}_{2N}\, \mbox{Tr} X_{11}
\end{array} \right) \ ,
\end{equation}

\begin{Proposition} A map $\Phi^U_{4N}$ is positive, indecomposable and optimal.
\end{Proposition}

Finally, we generalize formula (\ref{100}): let $z_{ij} \in \mathbb{C}$ for $i \neq j$ such that $z_{ij} = z_{ji}^*$. One defines
\begin{eqnarray}  \label{100-z}
\fl
 \Phi^{({\bf z})}_{2N}\left(\begin{array}{c|c|c|c}
X_{11} & X_{12} & \cdots & X_{1N}\\
\hline X_{21} & X_{22} & \cdots & X_{2N}\\
\hline \vdots & \vdots & \ddots & \vdots\\
\hline X_{N1} & X_{N2} & \cdots & X_{NN}\end{array}\right)=\frac{1}{2(N-1)}\left(\begin{array}{c|c|c|c}
A_{1} &  z_{12} B_{12} & \cdots &  z_{1N}B_{1N}\\
\hline z_{21} B_{21} & A_{2} & \cdots & z_{2N} B_{2N}\\
\hline \vdots & \vdots & \ddots & \vdots\\
\hline z_{N1} B_{N1} & z_{N2}B_{N2} & \cdots & A_{N}\end{array}\right)\ ,\label{Phi-2k}\end{eqnarray}
 where $A_k$ and $B_{kl}$ are defined by (\ref{Ak}) and (\ref{Bkl}), respectively.

\begin{Proposition}[\cite{DC10}] A map $\Phi^{({\bf z})}_{2N}$ is

\begin{itemize}
\item positive iff $|z_{ij}| \leq 1$,

\item  indecomposable and optimal iff $|z_{ij}| = 1$.

\end{itemize}
\end{Proposition}
For other constructions see also \cite{JAFA1,Zwolak-rec}.

\section{Circulant structures, Wyel operators and Bell-diagonal entanglement witnesses}  \label{CIRCULANT}

Let $\{e_0,\ldots,e_{d-1}\}$ be an orthonormal basis in $d$-dimensional Hilbert space and let $\mathbf{S}$ be a unitary shift operator defined by
\begin{equation}\label{}
  \mathbf{S}e_k = e_{k+1} \ , \ \ \ {\rm mod}\ d \ .
\end{equation}
One introduces \cite{CIRCULANT,ART} a family of linear $d$-dimensional subspaces in $\mathbb{C}^d \otimes \mathbb{C}^d$:
\begin{equation}\label{}
  \Sigma_0 = {\rm span}_\mathbb{C}\{ e_0 \otimes e_0, \ldots, e_{d-1} \otimes e_{d-1} \}\ ,
\end{equation}
and
\begin{equation}\label{}
  \Sigma_k = (\mathbb{I}_d \otimes \mathbf{S}^k) \Sigma_0\ , \ \ \ k=1,\ldots,d-1\ .
\end{equation}
Note, that $\Sigma_k$ and $\Sigma_l$ are mutually orthogonal (for $k \neq l$) and
\begin{equation}\label{}
  \Sigma_0 \oplus \Sigma_1 \oplus \ldots \oplus \Sigma_{d-1} = \mathbb{C}^d \otimes \mathbb{C}^d\ ,
\end{equation}
that is, a family $\Sigma_k$ provides a direct sum decomposition of $\mathbb{C}^d \otimes \mathbb{C}^d$.

\begin{Definition} We call an operator $A \in M_d(\mathbb{C}) \otimes M_d(\mathbb{C})$ circulant if
\begin{equation}\label{}
  A = A_0 \oplus A_1 \oplus \ldots \oplus A_{d-1} \ ,
\end{equation}
and $A_n$ is supported on $\Sigma_n$.
\end{Definition}
One has therefore
\begin{eqnarray}\label{}
A_n =  \sum_{i,j=0}^{d-1}\, a^{(n)}_{ij}\, E_{ij} \otimes {\bf S}^n\,
E_{ij}\, {\bf S}^{\dagger n} \ = \  \sum_{i,j=0}^{d-1}\, a^{(n)}_{ij}\, E_{ij}
\otimes E_{i+n,j+n}\ ,
\end{eqnarray}
and $a^{(n)}_{ij}$ is a collection of $d \times d$ complex matrices.
The crucial property of circulant states
is based on the following observation \cite{CIRCULANT,ART}: the
partially transposed circulant state $\rho$ displays similar
circulant structure, that is,
\begin{equation}\label{}
    ({\rm id} \otimes {\rm T}) A  = \widetilde{A}_0 \oplus \ldots
    \oplus \widetilde{A}_{d-1} \ ,
\end{equation}
where the operators $\widetilde{A}_n$ are supported on the new
collection of subspaces $\widetilde{\Sigma}_n$ which are defined as
follows:
\begin{equation}\label{}
    \widetilde{\Sigma}_0 = {\rm span}\{ e_0 \otimes e_{\pi(0)}, e_1 \otimes e_{\pi(1)}, \ldots, e_{d-1} \otimes e_{\pi(d-1)}
    \} \ ,
\end{equation}
where $\pi$ is a permutation defined by $\pi(k) = -k\ ({\rm mod} \ d)$.
The remaining subspaces $\widetilde{\Sigma}_n$ are defined by a
cyclic shift
\begin{equation}\label{}
    \widetilde{\Sigma}_n = (\mathbb{I} \otimes {\bf S}^n) \widetilde{\Sigma}_0 \ , \ \ n=1,\ldots,d-1\
    .
\end{equation}
Again, the collection $\{ \widetilde{\Sigma}_0, \ldots ,
\widetilde{\Sigma}_{d-1}\}$ defines direct sum decomposition of
$\mathbb{C}^d \otimes \mathbb{C}^d$
\begin{equation}\label{}
 \widetilde{\Sigma}_0 \oplus \ldots
\oplus \widetilde{\Sigma}_{d-1} = \mathbb{C}^d \otimes \mathbb{C}^d  \ .
\end{equation}
Moreover, operators $\widetilde{A}_n$ satisfy \cite{CIRCULANT}
\begin{eqnarray}\label{}
\fl
 \widetilde{A}_n = \sum_{i,j=0}^{d-1}\, \widetilde{a}^{(n)}_{ij}\,
E_{ij} \otimes {\bf S}^n\, E_{\pi(i)\pi(j)}\, {\bf S}^{\dagger n} \ = \
\sum_{i,j=0}^{d-1}\, \widetilde{a}^{(n)}_{ij}\, E_{ij} \ot
E_{\pi(i)+n,\pi(j)+n}\ ,
\end{eqnarray}
with
\begin{equation}\label{a-tilde}
\widetilde{a}^{(n)} \, =\, \sum_{m=0}^{d-1}\, a^{(n+m)} \circ (\Pi\,
{\bf S}^m)\ , \ \ \ \ \ \ \ \ (\mbox{mod $d$})\ ,
\end{equation}
where $\Pi$ is a permutation matrix corresponding to $\pi$, that is, $\Pi_{kl} = \delta_{k,\pi(l)}$, and $A \circ B$ denotes the Hadamard product of
matrices $A$ and $B$.

\begin{Example}
For $d=2$ one finds
\begin{eqnarray*}
% \nonumber to remove numbering (before each equation)
  \Sigma_0 &=& {\rm span}_\mathbb{C}\{ e_0 \otimes e_0, e_{1} \otimes e_{1} \}\ ,  \\
  \Sigma_1 &=& {\rm span}_\mathbb{C}\{ e_0 \otimes e_1, e_{1} \otimes e_{0} \}\ ,
\end{eqnarray*}
and $\widetilde{\Sigma}_0=\Sigma_1$,  $\widetilde{\Sigma}_1 = \Sigma_0$.
Hence a circulant operator $A$ and its partial transpose $A^\Gamma$ read
\begin{equation}\label{2C}
    A = \left( \begin{array}{cc|cc}
    a_{00} & \cdot & \cdot & a_{01} \\
    \cdot      & b_{00} & b_{01} & \cdot \\ \hline
    \cdot      & b_{10} & b_{11} & \cdot \\
    a_{10} & \cdot & \cdot & a_{11} \end{array} \right)\ , \ \ \
    A^\Gamma = \left( \begin{array}{cc|cc}
    \widetilde{a}_{00} & \cdot & \cdot & \widetilde{a}_{01} \\
    \cdot      & \widetilde{b}_{00} & \widetilde{b}_{01} & \cdot \\ \hline
    \cdot      & \widetilde{b}_{10} & \widetilde{b}_{11} & \cdot \\
    \widetilde{a}_{10} & \cdot & \cdot & \widetilde{a}_{11} \end{array} \right)\
    ,
\end{equation}
where the matrices $\widetilde{a} = [\widetilde{a}_{ij}]$ and
$\widetilde{b} = [\widetilde{b}_{ij}]$ read as follows
\begin{equation}\label{}
    \widetilde{a} = \left( \begin{array}{cc}
    a_{00} & b_{01} \\
    b_{10} & a_{11} \end{array} \right) \ , \ \ \ \
\widetilde{b} = \left( \begin{array}{cc}
    b_{00} & a_{01} \\
    a_{10} & b_{11} \end{array} \right)\ .
\end{equation}
Actually, circulant two qubit operators are very popular in the literature. For example in quantum optics  a circulant two qubit state is called an X-state. Hence, circulant states in $\mathbb{C}^d \otimes \mathbb{C}^d$ provide a natural generalization of X-states.
\end{Example}

\begin{Example}
For $d=3$ one finds
\begin{eqnarray*}
% \nonumber to remove numbering (before each equation)
\fl
  \Sigma_0 &=& {\rm span}_\mathbb{C}\{ e_0 \otimes e_0, e_{1} \otimes e_{1}, e_2 \otimes e_2 \}\ , \ \  \ \ \   \widetilde{\Sigma}_0 = {\rm span}_\mathbb{C}\{ e_0 \otimes e_0, e_{1} \otimes e_{2}, e_2 \otimes e_1 \}\ ,  \\
\fl
  \Sigma_1 &=& {\rm span}_\mathbb{C}\{ e_0 \otimes e_1, e_{1} \otimes e_{2}, e_2 \otimes e_0\}\ , \ \ \  \ \  \widetilde{\Sigma}_1 = {\rm span}_\mathbb{C}\{ e_0 \otimes e_1, e_{1} \otimes e_{0}, e_2 \otimes e_2\}\ , \\
\fl
  \Sigma_2 &=& {\rm span}_\mathbb{C}\{ e_0 \otimes e_2, e_{1} \otimes e_{0}, e_2 \otimes e_1\}\ , \ \ \ \ \  \widetilde{\Sigma}_2 = {\rm span}_\mathbb{C}\{ e_0 \otimes e_2, e_{1} \otimes e_{1}, e_2 \otimes e_0\}\ .
\end{eqnarray*}
%and
%\begin{eqnarray*}
% \nonumber to remove numbering (before each equation)
%  \widetilde{\Sigma}_0 &=& {\rm span}_\mathbb{C}\{ e_0 \otimes e_0, e_{1} \otimes e_{2}, e_2 \otimes e_1 \}\ ,  \\
%  \widetilde{\Sigma}_1 &=& {\rm span}_\mathbb{C}\{ e_0 \otimes e_1, e_{1} \otimes e_{0}, e_2 \otimes e_2\}\ ,\\
%  \widetilde{\Sigma}_2 &=& {\rm span}_\mathbb{C}\{ e_0 \otimes e_2, e_{1} \otimes e_{1}, e_2 \otimes e_0\}\ .
%\end{eqnarray*}
Hence a circulant two qutrit operator has the following structure

\begin{equation*}\label{3C}
%A =
\fl
  A = \left( \begin{array}{ccc|ccc|ccc}
    a & \cdot & \cdot & \cdot & a & \cdot & \cdot & \cdot & a \\
    \cdot& b & \cdot & \cdot & \cdot& b & b & \cdot & \cdot  \\
    \cdot& \cdot & c & c & \cdot & \cdot & \cdot & c &\cdot   \\ \hline
    \cdot & \cdot & c & c & \cdot & \cdot & \cdot & c & \cdot \\
    a & \cdot & \cdot & \cdot & a & \cdot & \cdot & \cdot & a  \\
    \cdot& b & \cdot & \cdot & \cdot & b & b & \cdot & \cdot  \\ \hline
    \cdot & b & \cdot & \cdot& \cdot & b & b & \cdot & \cdot \\
    \cdot& \cdot & c & c & \cdot& \cdot & \cdot & c & \cdot  \\
    a & \cdot& \cdot & \cdot & a & \cdot& \cdot & \cdot & a
     \end{array} \right)\ , \ \ \ \  A^\Gamma =
     \left( \begin{array}{ccc|ccc|ccc}
    x & \cdot & \cdot & \cdot &  \cdot & x & \cdot &  x & \cdot  \\
    \cdot& y & \cdot & y & \cdot & \cdot&   \cdot & \cdot & y  \\
    \cdot& \cdot& z &  \cdot & z & \cdot &  z & \cdot & \cdot   \\ \hline
    \cdot & y & \cdot &  y & \cdot & \cdot & \cdot  & \cdot & y \\
     \cdot & \cdot & z & \cdot & z & \cdot & z & \cdot & \cdot   \\
    x & \cdot & \cdot & \cdot & \cdot & x & \cdot & x & \cdot  \\ \hline
    \cdot &  \cdot & z & \cdot & z & \cdot &  z & \cdot & \cdot \\
    x & \cdot & \cdot & \cdot & \cdot & x & \cdot & x & \cdot  \\
     \cdot & y & \cdot & y & \cdot &  \cdot& \cdot & \cdot & y
     \end{array} \right)\ ,
\end{equation*}
where we schematically denote by `$a$' matrix elements supported on $\Sigma_0$, by `$b$' and `$c$' matrix elements supported on $\Sigma_1$ and $\Sigma_2$, respectively. Similarly, for a partially transposed operator `$x$' stands for a matrix elements supported on $\widetilde{\Sigma}_0$, by `$y$' and `$z$'  matrix elements supported on $\widetilde{\Sigma}_1$ and $\widetilde{\Sigma}_2$, respectively.

\fl
%\widetilde{a} &=& \left( \begin{array}{ccc}
%    a_{00} & c_{01} & b_{02} \\
%    c_{10} & b_{11} & a_{12} \\
%    b_{20} & a_{21} & c_{22} \end{array} \right) \ , \ \ \
%\widetilde{b} \ = \left( \begin{array}{ccc}
%    b_{00} & a_{01} & c_{02} \\
%    a_{10} & c_{11} & b_{12} \\
%    c_{20} & b_{21} & a_{22} \end{array} \right)\ ,
%\ \ \
%\widetilde{c} \ = \ \left( \begin{array}{ccc}
%    c_{00} & b_{01} & a_{02} \\
%    b_{10} & a_{11} & c_{12} \\
%    a_{20} & c_{21} & b_{22} \end{array} \right)\ .
%\end{eqnarray*}

\end{Example}
Now, let us define a collection of unitary Weyl operators \begin{equation}\label{U_mn}
    U_{mn} e_k = \lambda^{mk} \mathbf{S}^n  e_k = \lambda^{mk} e_{k+n}\ , \ \ \ \ {\rm mod}\ d \ ,
\end{equation}
with
%\begin{equation}\label{}
 $   \lambda= e^{2\pi i/d}$.
%\end{equation}
The matrices $U_{mn}$ satisfy
\begin{equation}\label{}
  U_{mn} U_{rs} = \lambda^{ms} U_{m+r,n+s} \ , \ \ \ U_{mn}^\dagger = \lambda^{mn} U_{-m,-n}\ ,
\end{equation}
and the following orthogonality relations
\begin{equation}\label{}
    {\rm tr}(U_{mn} U_{rs}^\dagger) = d\, \delta_{mr} \delta_{ns} \ .
\end{equation}
Some authors \cite{Pit} call $U_{mn}$ generalized spin
matrices since for $d=2$ they reproduce standard Pauli matrices:
\begin{equation}\label{U-sigma}
    U_{00} = \mathbb{I}\ , \ U_{01} = \sigma_1\ , \ U_{10} = i
    \sigma _2\ , \ U_{11} = \sigma_3\ .
\end{equation}
Weyl operators $U_{kl}$ in $\mathbb{C}^d$ may be used to construct circulant operators in $\mathbb{C}^d \otimes \mathbb{C}^d$.
 It is easy to show that
\begin{equation}\label{AUU}
  A = \sum_{k,l=0}^{d-1} c_{kl}\, U_{kl} \otimes U_{-kl}\ ,
\end{equation}
with $c_{kl} \in \mathbb{C}$ defines a  circulant operator. Clearly, the converse is not true, i.e. there are circulant operators which can not be represented via (\ref{AUU}).
Interestingly, one has the following
\begin{Proposition}[\cite{Bertlman2}] Let $W$ be a Hermitian circulant operator defined by
\begin{equation}\label{}
  W = a \left( (d-1) \mathbb{I}_d \otimes \mathbb{I}_d + \sum_{k,l=0;\, k+l > 0}^{d-1} c_{kl}\, U_{kl} \otimes U_{-kl} \right)\ ,
\end{equation}
with complex $c_{kl}$ and $a > 0$. If $|c_{kl}| \leq 1$, then $W$ is block positive.
\end{Proposition}

\begin{Example} If $d=2$ one finds
\begin{equation}\label{W2UU}
 W = a \left( \begin{array}{cc|cc} 1+\gamma & . & . & \alpha+\beta \\ . & 1-\gamma & \alpha-\beta & . \\ \hline
 . & \alpha - \beta & 1-\gamma & . \\ \alpha+\beta & . & . & 1+\gamma \end{array} \right) \ ,
\end{equation}
which is block positive whenever $a>0$ and $|\alpha|, |\beta|, |\gamma| \leq 1$.
\end{Example}
Now, let us define so called generalized Bell states in $\mathbb{C}^d \otimes \mathbb{C}^d$
\begin{equation}\label{}
  |\psi_{mn}\> = \mathbb{I}_d \otimes U_{mn} |\psi^+_d\> \ .
\end{equation}
One easily cheques $\< \psi_{mn}|\psi_{kl}\> = \delta_{mk}\delta_{nl}$. For $d=2$ they reproduce the standard two qubit Bell states
\begin{eqnarray*}
% \nonumber to remove numbering (before each equation)
  |\psi_{00}\>  &=& \frac{1}{\sqrt{2}} ( e_0 \otimes e_0 + e_1 \otimes e_1 ) \ , \ \ \ \   |\psi_{01}\> =  \frac{1}{\sqrt{2}} ( e_0 \otimes e_1 + e_1 \otimes e_0 ) \ ,  \\
  |\psi_{10}\> &=&  \frac{1}{\sqrt{2}} ( e_0 \otimes e_0 - e_1 \otimes e_1 )\ , \ \ \ \
  |\psi_{11}\>  =  \frac{1}{\sqrt{2}} ( e_0 \otimes e_1 - e_1 \otimes e_0 )\ .
\end{eqnarray*}
Finally, let us introduce a family of rank-1 projectors
\begin{equation}\label{}
  P_{mn} = |\psi_{mn}\>\<\psi_{mn}|\ .
\end{equation}
Let us observe that  $P_{mn}$ is supported on $\Sigma_n$ and
\begin{equation}\label{Pi_n}
    \Pi_n = P_{0n} + \ldots + P_{d-1,n} \ ,
\end{equation}
defines a projector onto $\Sigma_n$, i.e. $\,\Sigma_n = \Pi_n ( \mathbb{C}^d \otimes \mathbb{C}^d)$.

\begin{Cor} For an arbitrary operator  $A \in M_d(\mathbb{C}) \otimes M_d(\mathbb{C})$ the following projection
\begin{equation}\label{}
  \mathcal{P}(A) = \sum_{k=0}^{d-1} \Pi_k \, A\, \Pi_k\ ,
\end{equation}
defines a circulant operator.
\end{Cor}

\begin{Definition} A circulant operator $A \in M_d(\mathbb{C}) \otimes M_d(\mathbb{C})$ is called Bell diagonal if
\begin{equation}\label{}
  A = \sum_{k,l=0}^{d-1} a_{kl} P_{kl} \ ,
\end{equation}
that is, it is diagonal in the basis of generalized Bell diagonal states.
\end{Definition}

\begin{Proposition}
A circulant operator defined by (\ref{AUU}) is Bell diagonal.
\end{Proposition}

\begin{Example} Consider once more $W$ defined by (\ref{W2UU}). Note that for $\alpha=\beta=1$, $\gamma=0$ and $a=\frac 12$ one finds $W = \mathbb{F}$, i.e. one reconstructs a flip operator in $\mathbb{C}^2 \otimes \mathbb{C}^2$. Note, that
\begin{equation}\label{Flip-PPPP}
  \mathbb{F} = P_{00} + P_{01} + P_{10} - P_{11}\ ,
\end{equation}
which proves that $\mathbb{F}$ is Bell diagonal with a single negative eigenvalue.
\end{Example}
Actually, several entanglement witnesses considered so far are Bell diagonal.

\begin{Example} One finds for $W[a,b,c]$ defined in (\ref{W-abc})
\begin{eqnarray}\label{W-abc-P}
W[a,b,c] = (a-2)P_{00} + (a+1)(P_{10} + P_{20}) + b\Pi_{1} + c
\Pi_{2} \ ,
\end{eqnarray}
which shows that $W[a,b,c]$ is Bell diagonal with a single negative
eigenvalue `$a-2$'. Entanglement witness corresponding to the reduction map
$R_d(X) = \mathbb{I}_d {\rm tr}X - X$ in $M_d(\mathbb{C})$ reads as follows
\begin{eqnarray}\label{W-b}
    W =  \mathbb{I}_d \otimes \mathbb{I}_d - dP^+_d  =
  \sum_{k,l=0}^{d-1} P_{kl} - dP_{00}  \ ,
\end{eqnarray}
which shows that $W$ is Bell diagonal with a single negative
eigenvalue $1-d$.
Finally, $W_{d,k}$ defined in (\ref{Wdk}) may be  represented as follows
\begin{equation}\label{}
     W_{d,k} = (d+1-k)\Pi_0 + \sum_{\ell=1}^{k-1} \Pi_\ell - dP_{00}\ ,
\end{equation}
showing that $W_{d,k}$ is Bell diagonal and the single negative
eigenvalue `$1-k$' corresponds to the maximally entangled state $P_{00}$.
Note that $W_{d,d} = \sum_{\ell=0}^{d-1} \Pi_\ell - dP_{00}$ reproduces (\ref{W-b}).
\end{Example}
For more examples see e.g. \cite{Hiesmayr1,Bertlman,Bertlman1,Bertlman2,Bertlman3,Krammer,TAKA}.

\section{Construction of $k$-Schmidt witnesses}   \label{K}

Block positive operators, contrary to positive ones, are not characterized by their spectra. In this section we provide a class of entanglement witnesses which are fully characterized by spectral properties, that is, the properties of eigenvalues and corresponding eigenvectors. Note, that any Hermitian operator may written as
\begin{equation}\label{}
  W = W_+ - W_- \ ,
\end{equation}
where $W_+ > 0$ and $W_- \geq 0$, i.e. $W_+$ is strictly positive and all zero-modes, if any, are incorporated into $W_-$. This simple observation enables one to perform the following construction \cite{CMP,weak}: let $\psi_\alpha$ ($\alpha
=1,\ldots,D=d_Ad_B$) be an orthonormal basis in $\mathcal{H}_A \ot
\mathcal{H}_B$ and denote by $P_\alpha$ the corresponding projector
$P_\alpha = |\psi_\alpha\>\<\psi_\alpha|$. It leads therefore to the
following spectral resolution of identity
\begin{equation}\label{}
\mathbb{I}_A \otimes \mathbb{I}_B = \sum_{\alpha=1}^D P_\alpha\ .
\end{equation}
Now, take $D$  semi-positive numbers $\lambda_\alpha \geq 0$ such
that $\lambda_\alpha$ is strictly positive for $\alpha > L$, and
define
\begin{equation}\label{}
    W_- = \sum_{\alpha=1}^L \lambda_\alpha P_\alpha\ , \ \ \ \
W_+ = \sum_{\alpha=L+1}^D \lambda_\alpha P_\alpha\ ,
\end{equation}
where $L$ is an arbitrary integer $0<L<D$. This construction
guarantees that $W_+$ is strictly positive and all zero modes and
strictly negative eigenvalues of $W$ are incorporated into $W_-$.
Consider normalized vector $\psi \in \mathcal{H}_A \ot
\mathcal{H}_B$ and let
\[ s_1(\psi) \geq \ldots \geq s_d(\psi)  \ ,\]
denote its Schmidt coefficients $(d=\min\{d_A,d_B\})$. For any $1
\leq k \leq d$ one defines  $k$-norm of $\psi$ by the following
formula
\begin{equation}\label{}
    || \psi ||^2_k = \sum_{j=1}^k s^2_j(\psi)\ .
\end{equation}
It is clear that
\begin{equation}\label{}
    ||\psi ||_1 \leq ||\psi||_2 \leq \ldots \leq ||\psi ||_d \ .
\end{equation}
Note that $||\psi||_1$ gives the maximal Schmidt coefficient of
$\psi$, whereas due to the normalization, $||\psi||^2_d =
||\psi||^2 =1$. In particular, if $\psi$ is maximally entangled
then $s_1(\psi) = \ldots = s_d(\psi) = \frac 1d$ and hence
%\begin{equation}\label{}
 $   ||\psi||^2_k = \frac{k}{d}$.
%\end{equation}
Equivalently one may define $k$-norm of $\psi$ by
\begin{equation}\label{}
    ||\psi||^2_k = \max_\phi |\<\psi|\phi\>|^2\ ,
\end{equation}
where the maximum runs over all normalized vectors $\phi \in \HA \otimes \HB$ such that
${\rm SR}(\phi) \leq k$. Now, for any integer $\ell \geq 1$ such that $1-\sum_{\alpha=1}^L  ||\psi_\alpha||^2_\ell > 0$
let us define
\begin{equation}  \label{mu-l}
    \mu_\ell = \frac{\sum_{\alpha=1}^L \lambda_\alpha
    ||\psi_\alpha||^2_\ell}{1-\sum_{\alpha=1}^L
    ||\psi_\alpha||^2_\ell}\ .
\end{equation}
It is clear that $\mu_{\ell -1} \leq \mu_\ell  $.

\begin{Theorem}[\cite{CMP}] \label{TH-K} Let $\sum_{\alpha=1}^L ||\psi_\alpha||^2_{k} < 1$. If
\begin{equation} \label{T1}
     \lambda_\alpha \geq \mu_k\ , \ \ \
    \alpha=L+1,\ldots,D\ ,
\end{equation}
then $W \in \mathfrak{L}_k$. If moreover $\sum_{\alpha=1}^L
||\psi_\alpha||^2_{k+1} < 1$ and
\begin{equation}\label{T2}
    \mu_{k+1} > \lambda_\alpha\ , \ \ \
    \alpha=L+1,\ldots,D\ ,
\end{equation}
then $W \notin \mathfrak{L}_{k+1}$, that is, $W$ is a  $(k+1)$-Schmidt witness.
\end{Theorem}
Interestingly, this simple construction recovers many well know
examples of EWs.

\begin{Remark}
If $d_1=d_2=d$ and $P_1$ is the maximally entangled state in
$\mathbb{C}^d \otimes \mathbb{C}^d$, then the above theorem reproduces old result by
Takasaki and Tomiyama \cite{TT}. For $d_1=d_2=d\,$, $k=1$ and arbitrary $P_1$ the formula
$\lambda_\alpha \geq \mu_1$  $(\alpha=2,\ldots,d^2)$ was derived Benatti et al  \cite{Fabio} (see also \cite{Gniewko-JPA}). Recently, this class of witnesses was further analyzed in \cite{Hu}.
\end{Remark}

\begin{Example} Flip operator in $d_A=d_B=2$. Using (\ref{Flip-PPPP}) one finds
$$  W_+ = P_{00} + P_{01} + P_{10} \ , \ \ \ \ W_- = P_{11} \ , $$
and $\lambda_1=\lambda_2=\lambda_3=\lambda_4=1$. Note, that all projectors $P_{ij}$ are maximally entangled and hence one easily finds $\mu_1=1$ which shows that  condition (\ref{T1}) is trivially
satisfied $\lambda_\alpha \geq \mu_1$ for $\alpha =2,3,4$. We stress
that our construction does not recover flip operator in $d>2$. It
has $d(d-1)/2$ negative eigenvalues. Our construction leads to at
most $d-1$ negative eigenvalues.
\end{Example}

\begin{Example} For an entanglement witness (\ref{W-b}) corresponding to the reduction
map one has
\begin{equation}\label{+-}
     W_+ = \mathbb{I}_d \otimes \mathbb{I}_d -  P^+_d\ , \ \ \ W_- =  \lambda_1 P^+_d\ ,
\end{equation}
with $\lambda_1 = d-1$ and $\lambda_2 = \ldots = \lambda_D = 1$.  Again, one finds $\mu_1=1$ and hence condition
(\ref{T1}) is trivially satisfied $\lambda_\alpha \geq \mu_1$ for
$\alpha =2,\ldots,D=d^2$. Now, since $\psi_1$ corresponds to the
maximally entangled state one has $1 - ||\psi_1||^2_2 = (d-2)/d<1$.
Hence, condition (\ref{T2})
\begin{equation}\label{}
    \mu_2 = 2 \frac{d-1}{d-2} > \lambda_\alpha \ , \ \ \alpha
    =2,\ldots,D\ ,
\end{equation}
implies that $W$ is a 2-Schmidt witness. Equivalently, it shows that the reduction map $R_d$ is not 2-positive.
\end{Example}

\begin{Example} A family of witnesses in $\mathbb{C}^d \ot
\mathbb{C}^d$ defined by (\ref{+-})
\[ \lambda_1 = pd-1, \ \ \lambda_2 = \ldots = \lambda_D = 1  \ , \]
with $p\geq 1$ (see \cite{Pawel}).
Clearly, for $p=1$ it reproduces (\ref{W-b}). Now, conditions
(\ref{T1}) and (\ref{T2}) imply that if
\begin{equation}\label{kk}
    \frac{1}{k+1} < p \leq \frac{1}{k}\ ,
\end{equation}
then $W$ is a $k$-Schmidt witness.
Note, that if $p=1/d$, then $W=W_+ > 0$.
\end{Example}
Equivalently, we proved the following
\begin{Cor}
A linear map $\Phi : M_d(\mathbb{C}) \rightarrow M_d(\mathbb{C})$ defined by
\begin{equation}\label{Phi-p}
  \Phi_p(X) = \mathbb{I}_d \tr X - p X\ ,
\end{equation}
is $k$-positive but not $(k+1)$-positive if (\ref{kk}) is satisfied.
\end{Cor}

\begin{Example} A family of entanglement witnesses $W[a,b,c]$ defined  (\ref{W-abc}). Using (\ref{W-abc-P}) one finds
\[  \lambda_1 = 2-a\ ,\  \ \lambda_2=\lambda_3 = a+1 \ , \ \  \lambda_4=\lambda_5=\lambda_6=b \ , \ \  \lambda_7=\lambda_8=\lambda_9=c\ , \]
and hence $\mu_1 = (2-a)/2 $. Now, the condition (\ref{T1}) implies
\begin{enumerate}
\item $0 \leq a < 2\ $,
\item $ b,c \geq (2-a)/2\ $ .
\end{enumerate}
Note, that due to (\ref{ind}) all EWs satisfying the above conditions are decomposable.  Similarly one can check when
$W[a,b,c]$ belongs to $\mathfrak{L}_2$. One finds $\mu_2 = 2(2-a)$ and hence
condition (\ref{T1}) implies
\begin{enumerate}
\item $1 \leq a < 2\ $,
\item $ b,c \geq 2(2-a)\ $ .
\end{enumerate}
Note, that if $a+b+c=2$, then $W[a,b,c] \notin \mathfrak{L}_2$ and hence it provides a 2-Schmidt witness.
\end{Example}
One may ask a natural question: are witnesses constructed this way indecomposable or decomposable? The answer is provided by the following
\begin{Proposition}[\cite{weak}] An entanglement witness satisfying Theorem \ref{TH-K} is decomposable.
\end{Proposition}
Hence, these witnesses can not be used to detected PPT entangled state. However, having a decomposable linear map which is $k$-positive ($k>2$) one may construct a map which is indecomposable

\begin{Proposition}[\cite{Piani}]
Let $\Lambda$ be a $k$-positive map. $\Lambda$ is completely positive if and only if $\,{\rm id}_k \otimes \Lambda$ is decomposable.
\end{Proposition}
Hence, if $\Lambda$ is $k$-positive but not $(k+1)$-positive, then $\,{\rm id}_k \otimes \Lambda$ is necessarily indecomposable.
Consider for example a map defined by (\ref{Phi-p}). If $p$ satisfies (\ref{kk}), then ${\rm id}_k \otimes \Phi_p$ provides a positive indecomposable map.
See also \cite{Cl,Cl-rev} for related discussion.

\section{Multipartite entanglement witnesses}  \label{S-MULTI}

\subsection{Multipartite entanglement}

Consider now a multipartite quantum system living in $\mathcal{H}_{\rm total} = \mathcal{H}_1 \otimes \ldots \otimes \mathcal{H}_N$, where $N$ denotes a number of parties  (or subsystems). A vector $\psi \in \HT$ is separable (or fully separable, or $N$-separable) if $\psi = \psi_1 \otimes \ldots \otimes \psi_N$ such that $\psi_k \in \mathcal{H}_k$. Similarly, a positive  operator $X \in \mathfrak{L}_+(\HT)$ is fully separable if
\begin{equation}\label{}
  X = \sum_k A_k^{(1)} \otimes \ldots \otimes A_k^{(N)}\ ,
\end{equation}
where $A_k^{(i)} \in \mathfrak{L}_+(\mathcal{H}_i)$ for $i=1,\ldots,N$. Note, that the multipartite case is much more subtle. Apart from the full separability one may have a partial separability.

\begin{Definition} A vector $\psi \in \HT$  is separable with respect to a given partition $\{I_1, \ldots, I_k\}$, where $I_i$ are disjoint subsets of the indices $I=\{1, \ldots, N\}$,  $\bigcup_{j=1}^k I_j = I$, if and only if $\psi \in  \mathcal{H}_{I_1} \otimes \ldots \otimes \mathcal{H}_{I_k}$ can be written
\begin{equation}\label{}
  \psi = \psi_{I_1} \otimes \ldots \otimes \psi_{I_k} \ .
\end{equation}
A vector is bi-separable if $k=2$ and semiseparable if it is bi-separable and $|I_1|=1$ or $|I_2|=1$.
\end{Definition}
A similar definition of partial separability applies for positive operators from $\mathfrak{L}_+(\HT)$.

\begin{Example}
Consider three qubit case with $\HT = \mathbb{C}^2 \otimes \mathbb{C}^2 \otimes \mathbb{C}^2$. For $N=3$ there are only two notions of separability: full 3-partite separability
\begin{equation}\label{}
  \psi = \psi_1 \otimes \psi_2 \otimes \psi_3\ ,
\end{equation}
and bi-separability
\begin{equation}\label{}
  \psi = \psi_1 \otimes \psi_{23}\ , \ \ \psi = \psi_2 \otimes \psi_{13}\ , \ \ \psi = \psi_{12} \otimes \psi_3\ .
\end{equation}
A three qubit vector $\psi$ is called  genuine entangled if it is neither 3-separable nor bi-separable. Examples of such vectors are provided by celebrated {\rm GHZ} state
\begin{equation}\label{GHZ}
  |{\rm GHZ}\> =  \frac{1}{\sqrt{2}} (e_0 \otimes e_0 \otimes e_0 + e_1 \otimes e_1 \otimes e_1) \ ,
\end{equation}
and W state
\begin{equation}\label{W}
  |{\rm W}\> =  \frac{1}{\sqrt{3}} (e_0 \otimes e_0 \otimes e_1 + e_0 \otimes e_1 \otimes e_0 +  e_1 \otimes e_0 \otimes e_0) \ ,
\end{equation}
where $\{e_0,e_1\}$ denotes an orthonormal basis in $\mathbb{C}^2$. A vector $\psi \in \HT$ is equivalent to GHZ if $\psi = [A_1 \otimes A_2 \otimes A_3] |{\rm GHZ}\>$ and $A_k$ are invertible $2\times 2$ matrices, and it is equivalent to W if $\psi = [A_1 \otimes A_2 \otimes A_3] |{\rm W}\>$.
Actually, there are four classes of  positive operators $X \in \mathfrak{L}_+(\mathbb{C}^2 \otimes \mathbb{C}^2 \otimes \mathbb{C}^2)$:
two (fully and partially) separable classes
\begin{itemize}

\item fully separable class: $\mathfrak{L}_{\rm full}  = \{ \ \sum_k |\psi_k\>\<\psi_k| \ |\ \psi_k \ \mbox{-- {\rm fully\ separable}} \}$,

\item bi-separable:  $\mathfrak{L}_{\rm bi-sep}  = \{ \ \sum_k |\psi_k\>\<\psi_k| \ |\ \psi_k \ \mbox{-- {\rm bi-separable}} \}$,

\end{itemize}
and two classes of genuine entangled operators

\begin{itemize}

\item W-class:  $\mathfrak{L}_{\rm W}  = \{ \ \sum_k |\psi_k\>\<\psi_k| \ |\ \psi_k \ \mbox{-- {\rm W-equivalent}} \} - \mathfrak{L}_{\rm bi-sep}$,

\item GHZ-class:  $\mathfrak{L}_{\rm GHZ}  = %
%\{ \ \sum_k |\psi_k\>\<\psi_k| \ |\ \psi_k \ \mbox{-- {\rm GHZ-equivalent}} \}
\mathfrak{L}_+(\mathbb{C}^3 \otimes \mathbb{C}^3) - \mathfrak{L}_{\rm W}$.
\end{itemize}

Note that $\mathfrak{L}_{\rm full} \subset \mathfrak{L}_{\rm bi-sep}$ and by construction $\mathfrak{L}_{\rm W}$ and $\mathfrak{L}_{\rm GHZ}$ are disjoint.
\end{Example}
It should be stressed that checking bipartite separability with respect to all bi-partitions is not enough to guarantee full separability.

The essential difference between bipartite and multipartite entanglement is that for $N>2$ in general there is  no analog of the Schmidt decomposition  for a vectors from  $\HT$ \cite{Peres-multi,Pati,Acin}. Note that GHZ state (\ref{GHZ}) admits Schmidt decomposition but W state (\ref{W}) does not.

\subsection{Entanglement witnesses}

It should be clear that contrary to the bipartite case there are several types of multipartite entanglement witnesses detecting different types of multipartite entanglement.
\begin{Definition}\label{III} An operator $W \in \mathfrak{L}(\HT)$ is called a multipartite entanglement witness for a partition $\{I_1,\ldots,I_k\}$ if and only if
\begin{equation}\label{}
  \< \psi_{I_1} \otimes \ldots \otimes \psi_{I_k}\, |\, W\, |\, \psi_{I_1} \otimes \ldots \otimes \psi_{I_k}\> \geq 0 \ ,
\end{equation}
for all $\psi_{I_\ell} \in \mathcal{H}_{I_\ell}$.
\end{Definition}
Note, that if $\tr (X W) < 0$, then $X$ cannot be represented as
\begin{equation}\label{}
  X = \sum_\alpha |\psi^\alpha_{I_1} \otimes \ldots \otimes \psi^\alpha_{I_k}\>\< \psi^\alpha_{I_1} \otimes \ldots \otimes \psi^\alpha_{I_k}| \ ,
\end{equation}
and hence it is not separable with respect to a partition $\{I_1,\ldots,I_k\}$. In particular taking $I_j = \{j\}$ one constructs a witness which detects all operators which are not fully separable, that is, $W$ satisfies
\begin{equation}\label{W-full}
  \< \psi_1 \otimes \ldots \otimes \psi_N\, |\, W\, |\, \psi_1 \otimes \ldots \otimes \psi_N\> \geq 0 \ ,
\end{equation}
for all $\psi_j \in \mathcal{H}_j$, $j=1,\ldots,N$.

\begin{Proposition}[\cite{Horodecki-n}] $W$ satisfies (\ref{W-full}) if and only if
\begin{equation}\label{}
  W = [{\rm id}_1 \otimes \Lambda]P^+_{11}\ ,
\end{equation}
where the linear map $\Lambda : \mathfrak{L}(\mathcal{H}_1) \rightarrow \mathfrak{L}(\mathcal{H}_2 \otimes \ldots \otimes \mathcal{H}_N)$ satisfies
\begin{equation}\label{}
  \Lambda^\#(A_2 \otimes \ldots \otimes A_N) \geq 0 \ ,
\end{equation}
for all $A_k \in \mathfrak{L}_+(\mathcal{H}_k)$. $P^+_{11}$ stands for a maximally entangled state in $\mathcal{H}_1 \otimes \mathcal{H}_1$.
\end{Proposition}
A linear map $\Phi :  \mathfrak{L}(\mathcal{H}_2 \otimes \ldots \otimes \mathcal{H}_N)  \rightarrow \mathfrak{L}(\mathcal{H}_1)$ which has to be positive but only on  products of positive operators  provides a natural generalization of a positive map.

\begin{Cor} $X$ is fully separable if and only if
\begin{equation}\label{}
  [{\rm id }_1 \otimes \Phi] X \geq 0 \ ,
\end{equation}
for all linear maps $\Phi :  \mathfrak{L}(\mathcal{H}_2 \otimes \ldots \otimes \mathcal{H}_N)  \rightarrow \mathfrak{L}(\mathcal{H}_1)$ which are positive on products of positive operators, that is, $\Phi(A_2 \otimes \ldots \otimes A_N) \geq 0$ for $A_k \in \mathfrak{L}_+(\mathcal{H}_k)$.
\end{Cor}

\begin{Example} Consider $\HT = \mathbb{C}^2 \otimes \mathbb{C}^2 \otimes \mathbb{C}^2$. One finds \cite{Guhne}
\begin{equation}\label{}
  W = \mathbb{I}_9 - \frac 32\, |{\rm W}\>\<{\rm W}|\ ,
\end{equation}
provides an entanglement witness of genuine entanglement, and
\begin{equation}\label{}
  W' = \mathbb{I}_9 - \frac 94\, |{\rm W}\>\<{\rm W}|\ ,
\end{equation}
may detect  states which are not fully separable. Finally,
\begin{equation}\label{}
  W'' = \mathbb{I}_9 - \frac 43\, |{\rm GHZ}\>\<{\rm GHZ}|\ ,
\end{equation}
detects states which are not in the W class.
\end{Example}

%\section{Convex cones - basic definitions}

\section{Geometric approach: convex cones, duality, extremality and optimality}  \label{S-CONES}

The basic problem we address in this paper is the characterization of a convex subset of block-positive operators within  a convex set of positive operators. Equivalently, such characterization provides the description of a convex subset of separable states within a convex set of all states of a composite quantum system. Interestingly, this is the special case of more general problem. In this section we provide a geometric approach to the general mathematical problem of identifying a convex subset of a given convex set.

\begin{Definition}
A closed subset $K$ of a linear space $X$ over $\mathbb{R}$ is called a convex cone if
for arbitrary   $x, y \in K$ and  $\lambda, \mu \in \mathbb{R}_+$ a convex combination $\lambda x + \mu y \in K$.
\end{Definition}
Now, if $B \subset X$ one denotes by $\mathrm{conv} B$ a minimal convex cone in $X$ containing $B$. Let $X^*$ be a dual space and denote by $\<y|x\>$ a natural pairing between $y\in X^*$ and $x \in X$. For any subset $B \subset X$ one defines
\begin{equation}\label{}
  B^\circ := \{\, y \in X^*\ |\ \forall x \in B \ \< y | x \> \ge 0 \,\} \ ,
\end{equation}
which is a convex cone in $X^*$. In particular if $K$ is a cone then $K^\circ$ is a dual cone in $X^*$. Note, that a double dual $B^{\circ\circ}$ defines a convex cone in $X$. In finite dimensional case we consider in this paper one has $\mathrm{conv} B = B^{\circ\circ}$ (this property still holds for infinite dimensional $X$ if one assumes that $X$ is reflexive).
The operation of taking a dual $B \rightarrow B^\circ$ enjoys the following properties: for any two subsets $B,C \subset X$
\begin{enumerate}

\item $ B \subset C \iff C^\circ \subset B^\circ$

\item $ (B \cap C)^\circ = \mathrm{conv}(B^\circ \cup C^\circ)$,

\item $(B \cup C)^\circ = B^\circ \cap C^\circ$.

\end{enumerate}
Let us list natural convex cones considered so far in this paper.

\begin{enumerate}

\item A convex cone  $\mathcal{L}_+(\mathcal{H})$ of positive operators in $\mathcal{H}$. This is a cone in $X = X^* = \mathbb{R}^{n^2}$, where $n={\rm dim}\, \mathcal{H}$. Note, that this cone is self-dual, that is, $\mathcal{L}_+(\mathcal{H})^\circ = \mathcal{L}_+(\mathcal{H})$ $(\mathcal{L}_+(\mathcal{H})$ is a cone of (unnormalized) states and $\mathcal{L}_+(\mathcal{H})^\circ$ is a cone of positive observables).

\item A covex cone $\mathfrak{L}_1 \subset \mathcal{L}_+(\HA  \otimes \HB)$ of separable operators. Its dual $\mathfrak{L}_1^\circ = \mathbb{W}_1$ provides a convex cone of block-positive operators  in $\HA \otimes \HB$. Clearly, $\mathcal{L}_+(\HA  \otimes \HB) \subset \mathbb{W}_1$.

\item A convex cone $\mathfrak{L}_k$ of positive operators $X$ in $\mathfrak{L}_+(\HAB)$ such that ${\rm SN}(X) \leq k \leq d=\min\{d_A,d_B\}$. Its dual $\mathfrak{L}_k^\circ = \mathbb{W}_k$ provides a convex cone of $k$-block-positive operators  in $\HA \otimes \HB$. One has
\begin{equation*}\label{}
  \mathfrak{L}_1 \subset \ldots \subset \mathfrak{L}_{d-1} \subset \mathfrak{L}_d = \mathfrak{L}_+(\HAB)\ ,
\end{equation*}
and by duality
\begin{equation*}\label{}
  \mathfrak{L}_+(\HAB) = \mathbb{W}_d \subset \mathbb{W}_{d-1} \subset \ldots \subset \mathbb{W}_{1} \ .
\end{equation*}

\item A convex cone $\mathfrak{L}_{\rm PPT} = \mathfrak{L}_+ \cap \mathfrak{L}_+^\Gamma$ of operators in $\HA \otimes \HB$. Its dual $\mathfrak{L}_{\rm PPT}^\circ = \mathbb{W}_{\rm DEC} = {\rm conv}(\mathfrak{L}_+ \cup \mathfrak{L}_+^\Gamma)$  provides a convex cone of decomposable operators, i.e. $A + B^\Gamma$, where $A,B \in \mathfrak{L}_+(\HAB)$. One has
\begin{equation}\label{}
  \mathfrak{L}_{\rm PPT} \subset  \mathfrak{L}_+(\HAB)\ , \ \ {\rm and} \ \  \mathfrak{L}_+(\HAB) \subset \mathbb{W}_{\rm DEC}\ .
\end{equation}

\item In the multipartite scenario a convex cone $\mathfrak{L}_{(I_1,\ldots ,I_k)} \subset \mathfrak{L}_+(\HT)$ of $(I_1,\ldots ,I_k)$-separable elements and $\mathbb{W}_{(I_1,\ldots ,I_k)} \supset \mathfrak{L}_+(\HT)$ of $(I_1,\ldots ,I_k)$-block-positive operators (cf. Definition \ref{III}).

\end{enumerate}

\begin{Definition}
If $K$ is a convex cone then a subset $F \subset K$ is called a face, iff any line segment contained in $K$ with inner point in $F$ is contained in $F$ (we write $F \triangleleft K$).
\end{Definition}
A face $F$ of a convex cone is again a convex cone. If $G \triangleleft F$, then one has $G \triangleleft F \triangleleft K$ and hence `$\triangleleft$' provides a partial order in the space of faces of $K$. The maximal element in this family is the cone $K$ itself and the minimal one is $\{0\}$. For each element $x \in K$ one may define a minimal face $F(x) \triangleleft K$ containing $x$. Now, if $F \triangleleft K$, then one defines a dual face
\begin{equation}\label{}
  %\Phi_K(F) =
  F' :=\{y \in K^\circ\ |\  \forall x \in F \ \ \braket{y}{x}=0 \} \ ,
\end{equation}
which is by construction a face in $K^\circ$. An operation $F \rightarrow F'$ defines an order reversing lattice isomorphism, i.e. if $F_1 \triangleleft F_2 \triangleleft K$ then  $F_2' \triangleleft F_1' \triangleleft K^\circ$.

\begin{Definition}
A face is exposed if $F = G'$ with $G \triangleleft K^\circ$, i.e. $F$ is a dual face.
\end{Definition}

\begin{Proposition}[\cite{conv}] A face $F \triangleleft K$ is exposed if and only if $F=F''$.
\end{Proposition}

% and one can consider its faces. Face of $F$ is also a face of $K$ so one has in a convex cone an partially ordered family of subcones (being a subspace %$\triangleleft$ is the relation of partial order). The maximal element in this family is the cone itself and the minimal one is $\{0\}$. Let us denote this %family as $\mathcal{F}_K$.

%  \paragraph{Duality of faces and exposedness} In the dual cone we have also an ordered structure of faces. One can define a map assigning to a given face of %$K$ a face of its dual cone $K^*$ by the formula: $\Phi_K(F) = \{y \in K^*: \forall x \in F \braket{y}{x}=0 \}$. This map is order reversing. In general, the %map $\Phi: \mathcal{F}_K \to \mathcal{F}_{K^*}$ is a monomorphism, and one must not produce all faces of $K^*$ in this way.

%  One has also a map acting in the opposite direction $\Phi_{K^*}: \mathcal{F}_{K^*} \to \mathcal{F}_K$. The sets $\Phi_{K^*} \circ \Phi_K (\mathcal{F}_K) %\subset \mathcal{F}_K$ and $\Phi_K \circ \Phi_{K^*} (\mathcal{F}_{K^*}) \subset \mathcal{F}_{K^*}$ may be proper subsets. The elements of these subsets are %called \textit{exposed faces}.

%  The maps $\Phi_K$ and $\Phi_{K^*}$ restricted to sets of exposed faces are isomorphisms (one beeing the inverse of the other) so the structures of exposed %faces of dual cones are isomorphic.

To illustrate the concept of duality consider the cones in $X,X^* \cong \mathbb{R}^3$. Note, that any cone in $\mathbb{R}^3$ may uniquely represented by its $2$-dimensional sections (for example $x_3=1$) which defines a convex set on $\mathbb{R}^2$. Let us observe that

%\begin{Example} \label{nonexp_faces}
%We will represent cones in  using their . First observe the following:
    \begin{enumerate}
     \item a unit disk $D$ in $X$ is self-dual. If we enlarge $D$ its dual $D^\circ$ decreases and vice versa,

     \item taking one point say $(x_0,y_0,1)$ then the dual cone is represented by a halfplane $(x,y,1)$ defined by $x x_0 + y y_0 + 1 \geq 0$.
    % halfplane (a cross-section of a halfspace). The closer the point is to zero, the further from zero is the halfplane. ????????
    \end{enumerate}

\begin{Example}  \label{nonexp_faces}
    Consider a cone $K \in X$ being represented by intersection of the unit disk and the halfplane  $x \leq \frac 12$.
    Using the above observations and the property of the duality operation one finds that the dual cone $K^\circ \subset X^*$ is represented by the convex hull of a unit circle and the point $(2,0,1)$.

    \begin{tabular}{ccc}
    \begin{minipage}{6.5cm}
    \includegraphics[width=6cm]{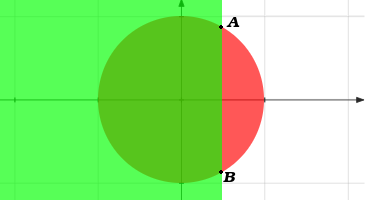}
    % K1.png: 365x200 pixel, 96dpi, 9.66x5.29 cm, bb=0 0 274 150
    Section of a cone $K$ is an intersection of a disk and a halfplane
    \end{minipage}
    \begin{minipage}{2cm}
    \quad`
    \end{minipage}
    \begin{minipage}{6.5cm}
    \includegraphics[width=6cm]{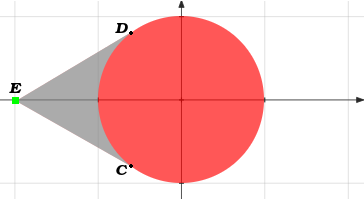}
    % K1.png: 365x200 pixel, 96dpi, 9.66x5.29 cm, bb=0 0 274 150
    Section of the dual cone $K^\circ$ is a convex hull of a disk and a point $\{E\}$.
    \end{minipage}
    \end{tabular}

    \vspace{1cm}

    The diagram below shows the structure of faces of $K$ and $K^\circ$:

 \begin{tabular}{cc}
 \begin{minipage}{7.5cm}
%  \begin{tabular}{cccccccl}
%   & & & & $K$ & & \\
%   & & & $\swarrow$ & & & & \\
%   & & $\overline{AB}$ & & & & & \\
%   & $\swarrow$ & & $\searrow$ & & & & \\
%   $A$ & & & & $B$ & & $\mathrm{Int}(\stackrel \frown {BA}$) \\
%   & $\searrow$ & & $\swarrow$ & & & & \\
%   & & $\{0\}$ & & & & &
%  \end{tabular}
 \begin{tikzpicture}[node distance=2cm, auto]
 \node (K) {$K$};
 \node (AB) [below of=K]{$\overline{AB}$};
 \node (A) [left of=AB, below of=AB] {$A$};
 \node (B) [right of=AB, below of=AB] {$B$};
 \node (I) [right of=B]{$\mathrm{Int}(\stackrel \frown {BA})$};
 \node (I1) [node distance=0.8cm, left of=I] {};
 \node (I2) [node distance=0.8cm, right of=I] {};
 \node (0) [right of=A, below of=A] {$\{0\}$};
 \draw[->] (K) to (AB);
 \draw[->] (AB) to (A);
 \draw[->] (AB) to (B);
 \draw[->] (A) to (0);
 \draw[->] (B) to (0);
 \draw[->, bend left] (K) to (I);
 \draw[->, bend left] (K) to (I1);
 \draw[->, bend left] (K) to (I2);
 \draw[->, bend left] (I1) to (0);
 \draw[->, bend left] (I2) to (0);
 \draw[->, bend left] (I) to (0);
\end{tikzpicture}
 \begin{center}
 Faces of $K$
 \end{center}
 \end{minipage}
 \begin{minipage}{7.5cm}
%  \begin{tabular}{ccccccccl}
%   & & $\{0\}$ & & & & & & \\
%   & $\textcolor{red}{\nearrow}$ & $\uparrow$ & $\textcolor{red}{\nwarrow}$ & & & & & \\
%   $\textcolor{red}{C}$ & & $E$ & & $\textcolor{red}{D}$ & & & & \\
%   $\textcolor{red}{\uparrow}$ & $\nearrow$ & & $\nwarrow$ & $\textcolor{red}{\uparrow}$ & & & & \\
%   $\overline{CE}$ & & & & $\overline{ED}$ & & & & $\mathrm{Int}(\stackrel \frown {DC}$) \\
%   & $\nwarrow$ & & $\nearrow$ & & & & & \\
%   & & $K^*$ & & & & & &
%  \end{tabular}
 \begin{tikzpicture}[node distance=2cm, auto]
 \node (K) {$\{0\}$};
 \node (AB) [below of=K]{$E$};
 \node (A) [left of=AB, below of=AB] {$\overline{CE}$};
 \node (B) [right of=AB, below of=AB] {$\overline{ED}$};
 \node (I) [right of=B]{$\mathrm{Int}(\stackrel \frown {DC})$};
 \node (I1) [node distance=0.8cm, left of=I] {};
 \node (I2) [node distance=0.8cm, right of=I] {};
 \node (0) [right of=A, below of=A] {$K^*$};
 \node (C) [left of=AB] {$\textcolor{red}{C}$};
 \node (D) [right of=AB] {$\textcolor{red}{D}$};
 \draw[->] (AB) to (K);
 \draw[->] (A) to (AB);
 \draw[->] (B) to (AB);
 \draw[->] (0) to (A);
 \draw[->] (0) to (B);
 \draw[->, bend right] (I) to (K);
 \draw[->, bend right] (I1) to (K);
 \draw[->, bend right] (I2) to (K);
 \draw[->, bend right] (0) to (I1);
 \draw[->, bend right] (0) to (I2);
 \draw[->, bend right] (0) to (I);
 \draw[red,->] (A) to (C);
 \draw[red,->] (B) to (D);
 \draw[red,->] (C) to (K);
 \draw[red,->] (D) to (K);
 \end{tikzpicture}
 \begin{center}
 Faces of $K^\circ$
 \end{center}
 \end{minipage}
 \end{tabular}
 \vspace{0.5cm}

If we remove from the diagram for $K^\circ$ the red part, then both diagrams are isomorphic via duality map. Red faces are non-exposed.
  \end{Example}

For the convex cones considered in this paper one has:

\begin{enumerate}

\item  The cone of positive operators, i.e. unnormalized states,  is dual to the cone of positive observables. These cones are isomorphic under the isomorphism between spaces of states and observables induced by the Hilbert-Schmidt inner product.
  A maximal face containing a density operator $\rho$ is the set of states whose image is contained in the image $V :=\Im(\rho)$ of $\rho$. The dual face is a set of observables with $\Im A \subset V^\perp$. It is easy to observe, that all faces of $\mathfrak{L}_+$ are exposed \cite{Kye-REV}. A state $\rho$ is an edge state if $F(\rho)$ does not contain a separable element.

\item  The lattice of faces on the cone of separable operators is poorly understood (see \cite{Alfsen,Kye-REV}).

\item   In the cone of PPT operators $\mathfrak{L}_{\rm PPT}$  the face is a set of PPT operators for which $\Im X \subset V_1, \Im X^\perp \subset V_2$, where $V_1, V_2$ are subspaces of the Hilbert space of the system \cite{Kye-REV}. However it is hard to determine which pairs of subspaces give rise to a face of the cone of PPT operators. Note, that $\rho$ defines an edge state if $F(\rho)$ does not contain a separable element.
    %A PPT state $\rho$ satisfies range criterion iff the smallest face containing $\rho$ contains a separable state in its interior \cite{facerange}.

\end{enumerate}
A convex cone is a sum of rays (a ray is a set of points in $K$ differing by a positive scalar).
\begin{Definition} A ray is extremal if it provides a 1-dimensional face.
\end{Definition}
A point from an extremal ray can not be represented as a convex combinations of points from other rays of the cone. Due to the  Straszewicz theorem \cite{conv} the exposed rays form a dense subset of extreme rays (in the topology in the set of rays induced from the standard topology of $X$).

\begin{Example}
   In the Example \ref{nonexp_faces} the extreme rays of $K$ are represented by points from closed arc $\mathrm{Int}(\stackrel \frown {BA})$ and all of them are exposed. The extreme points of $K^\circ$ are points of closed arc $\mathrm{Int}(\stackrel \frown {EC})$ and a point $D$. Points $E,C$ are not exposed.
\end{Example}

Consider a pair of convex cones $K \subset L$. The elements of $L \setminus K$ detect the elements of $K^\circ \setminus L^\circ$ in the sense, that for any element $X \in K^\circ \setminus L^\circ$ there exists an element $W \in L \setminus K$ such that $\braket{X}{W}<0$.
Conversely,  using the elements of $K^\circ \setminus L^\circ$ one can detect the elements of $K \subset L$.   For any element $W \in L \setminus K$ one defines the subset $\mathcal{D}_W$ of detected elements of $K^\circ \setminus L^\circ$. $W_1$ is finer than $W_2$ if $\mathcal{D}_{W_1} \supseteq \mathcal{D}_{W_2}$. Finally, $W$ is called \textit{optimal} if there is no other element in $L \setminus K$ which is finer than $W$.

\begin{Proposition}[\cite{opt,gtdo,cospan}] $W$ is optimal if and only if $\,F(W) \cap K = \{0\}$.
\end{Proposition}

\begin{figure}[!h]
%  \begin{minipage}[t]{5cm}
%  \vspace{0pt}
 \includegraphics[width=5cm]{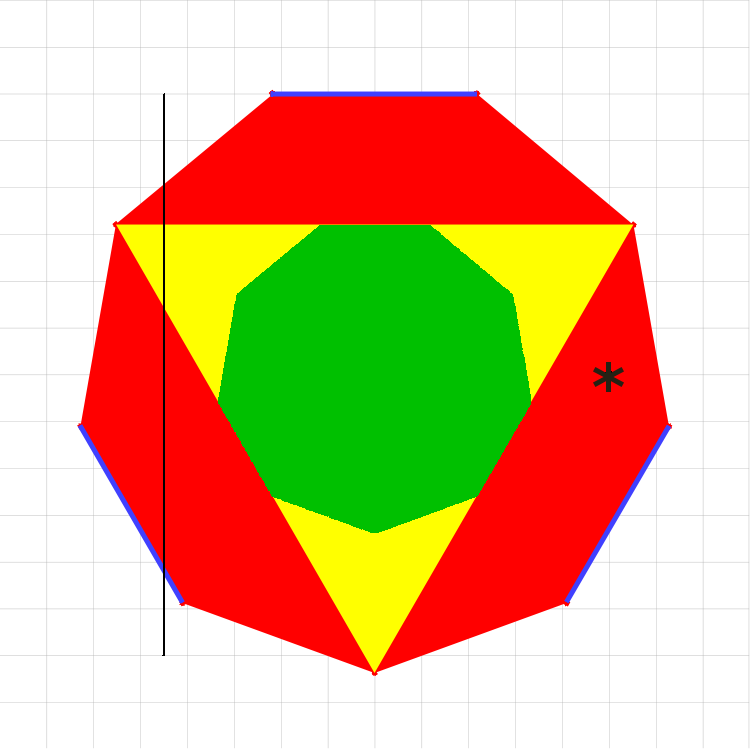}
 \includegraphics[width=5cm]{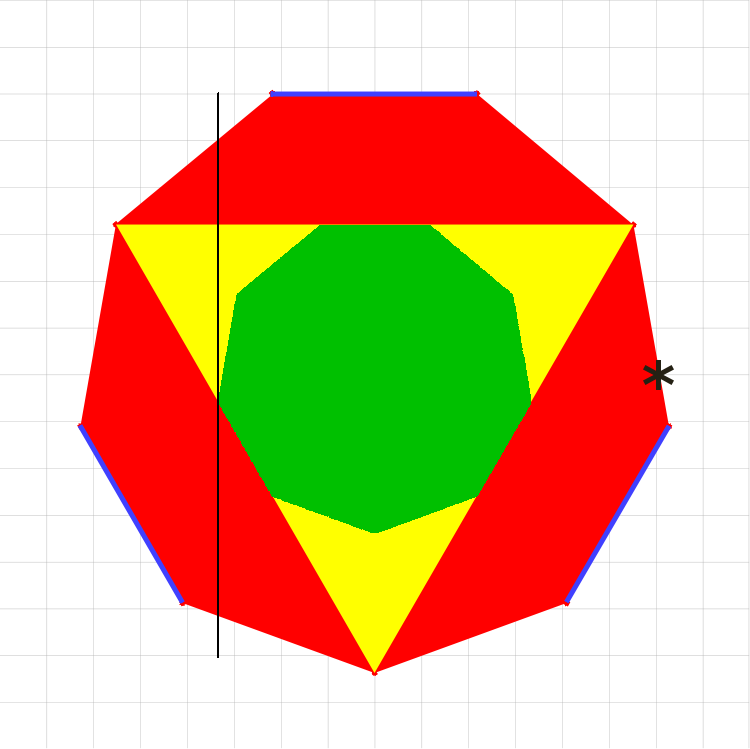}
 \includegraphics[width=5cm]{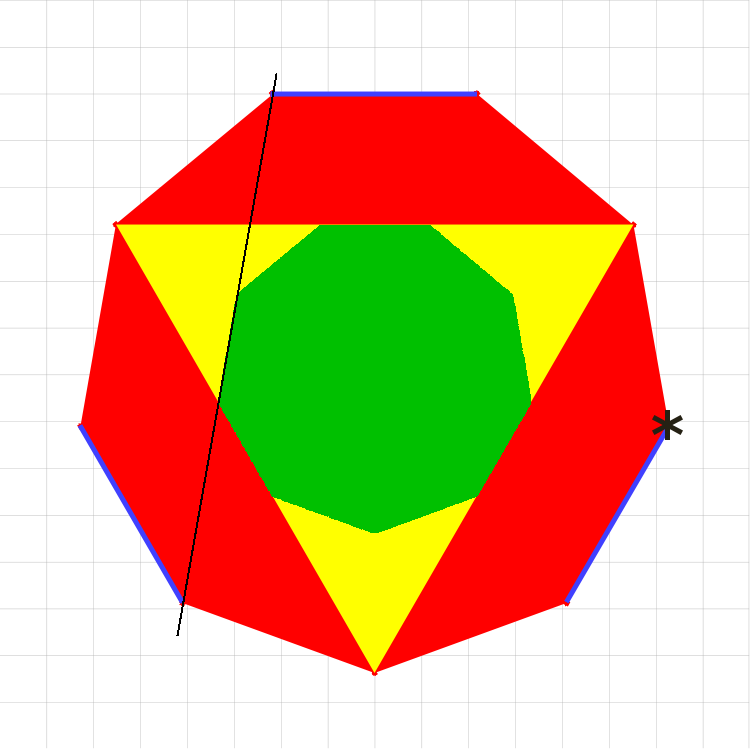}
 \caption{A convex cone  $K=K^\circ$ is represented by the yellow triangle. $L$ is represented by the red nonagone and $L^\circ$ by the green one. Left picture: the black element of $L\setminus K$ detect all elements on the right of the line. Middle picture: this element is finer than the previous one, but still not optimal - it can be `improved'. Right picture: this element is already optimal - cannot be further `improved'. The optimal witnesses are denoted  by blue segments.}
% }
\end{figure}
Recall, that an entanglement witness is optimal if it possesses a spanning property. Interestingly, this property may be reformulated as follows:  $[F(W)]' \cap \mathrm{Int}K^\circ \ne \emptyset$ (see \cite{gtdo}).  Actually, if all faces of $L$ are exposed, this condition is also  necessary for optimality.

 \begin{Example} [$K=\mathcal{L}_+ \subset L = \mathbb{W}_1$]
Consider elements of $\mathbb{W}_1 \setminus \mathcal{L}_+$ (entanglement witnesses) which detect elements of $\mathcal{L}_+ \setminus \mathcal{L}_1$ (entangled operators).    The face dual to $F(W)$ is spanned by  separable elements $| e \otimes f \rangle\langle e \otimes f |$ (extremal rays of $\mathcal{L}_1$) such that $\langle e \otimes f |W|e \otimes f \rangle = 0$. The interior rays of $\mathcal{L}_+$ are generated by positive matrices of full rank, equal to the dimension of the Hilbert space $\HAB$. A full rank matrix  belongs to the face $F(W)$ iff the set of vectors $\{ e \otimes f\ |\  \langle e \otimes f |W|e \otimes f \rangle = 0 \}$ spans the whole Hilbert space. Now, since there are nonexposed faces in $\mathbb{W}_1$, so the spanning condition is only sufficient but not necessary for optimality. An example of a witness which is optimal but does not possess a spanning property is the Choi witness -- its ray is extreme, but the smallest exposed face containing this witness contains also a positive matrix.
\end{Example}

\begin{Example}[$ K = \mathbb{W}_{\rm DEC}  \subset L=\mathbb{W}_1$]
Consider now elements of $\mathbb{W}_1 \setminus \mathbb{W}_{\rm DEC}$ (non-decomposable entanglement witnesses) which detect elements of $\mathcal{L}_{PPT} \setminus \mathcal{L}_1$ (PPT entangled operators). The face $F(W)$ is again spanned by pure separable states $| e \otimes f \rangle\langle e \otimes f |$ such that $\langle e \otimes f |W|e \otimes f \rangle = 0$, but the interior of $\mathcal{L}_{PPT}$ are states $\rho$ such that $\rho$ and $\rho^\Gamma$ are of full rank. An element of $F(W)$ is in $\mathrm{Int}(\mathcal{L}_{PPT})$ iff $\{ e \otimes f: \langle e \otimes f |W|e \otimes f \rangle = 0 \}$ spans the whole Hilbert space and $\{ e \otimes f \ |\  \langle e \otimes f^* |W|e \otimes f^* \rangle = 0 \}$ as well, so $W$ is nd-optimal if and only if $W$ is optimal and $W^\Gamma$ is optimal. We stress that there are non-decomposable witnesses which are optimal but not nd-optimal \cite{cospan}.
\end{Example}

\begin{figure}[h!]
\begin{center}
 \includegraphics[width=9cm]{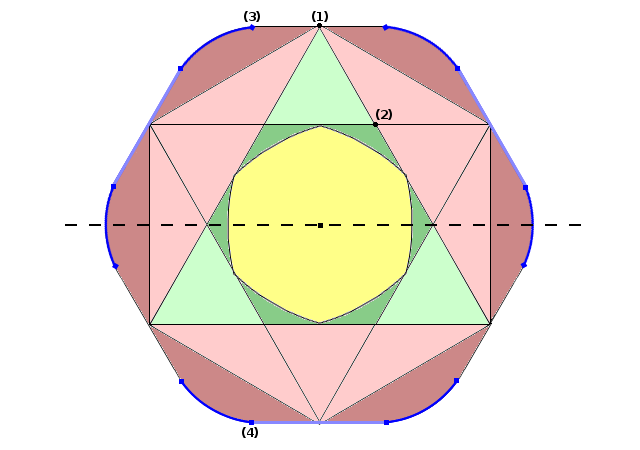}
\end{center}
 % wszc.png: 640x450 pixel, 72dpi, 22.57x15.87 cm, bb=0 0 640 450
 \caption{[Color online] Schematic representation  of states and entanglement witnesses}
 \label{F-LAST}
\end{figure}

We summarize our presentation in Fig. \ref{F-LAST}:  the green triangle denotes the self-dual cone of positive matrices.
 The dashed line is a set of fixed points of partial transposition –- partial transposition is reflection with respect to this line. The light-green regions are NPT entangled states.
 The light-red regions are decomposable EWs and point (1) is an example of an extremal decomposable EW.
 The dark green regions are PPT-entangled states, where point (2) is an example of an edge-state.
 The central yellow region denotes the separable states.
 The dark-red regions are indecomposable EWs.
 The blue boundary regions are optimal witnesses where the light-blue open segments denote the non-decomposable witnesses which are optimal, but not nd-optimal.
 A point (3) is an example of Choi-like witnesses, which are extremal hence optimal but not spanning (face generated by this element contains a positive matrix), hence not exposed. After partial transposition the point (3) is mapped into the point (4) which has a  spanning property (face generated by this element contains no positive elements).  Points (3) and (4) are nd-optimal without a bi-spanning property.

\section{Conclusions}

This paper provides a review of the theory of entanglement witnesses. From the physical point of view entanglement witnesses define a universal tool for analysis and classification of quantum entangled states. From the mathematical point of view they provide highly nontrivial generalization of positive operators and they find elegant correspondence with the theory of positive maps in matrix algebras (or more generally $\mathbb{C}^*$-algebras). We concentrate on theoretical analysis of various important notions like (in)decomposability, atomicity, optimality, extremality and exposedness. Several methods of construction are provided as well. Our discussion is illustrated by many examples enabling the reader to see the intricate structure of these objects. It is shown that the theory of entanglement witnesses finds elegant geometric formulation in terms of convex cones and related geometric structures.

It should be stressed that there are important topics not covered in this paper.
We basically concentrate on theoretical analysis.  For experimental realization of entanglement witnesses  see review paper by G\"uhne and Toth \cite{Guhne}. Recently, a new interesting topic {\em device independent approach to EWs} started to be discussed (see for example recent papers \cite{DEVICE1,DEVICE2}).

\section*{Acknowledgements}

This paper was partially supported by the National Science Center project  %{\em Quantum correlations: analysis, detection and dynamics}.
DEC-2011/03/B/ST2/00136. We thank Andrzej Kossakowski, Stanis{\l}aw L. Woronowicz, Remigiusz Augusiak, Jarek Korbicz, Maciej Lewenstein,  Seung-Hyeok Kye, Pawe{\l} Horodecki, Jacek Jurkowski, {\L}ukasz Skowronek, Adam Majewski, Justyna Zwolak, Ingemar Bengtsson, Marcin Marciniak and Adam Rutkowski for valuable discussions.

\end{document}